\documentclass[prd,nofootinbib,preprint,superscriptaddress]{revtex4-1}
\usepackage{graphicx}
\usepackage{amsmath}
\usepackage{bm}
\usepackage[compat=1.1.0]{tikz-feynman}
\usepackage{feynmp-auto}
\usepackage{tikzsymbols}
\usetikzlibrary{decorations.pathmorphing}
\usepackage{amsfonts}
\usepackage{amsmath}
\usepackage{amssymb}
\usepackage{physics}
\usepackage{tcolorbox}
\usepackage{circuitikz}
\usepackage{multirow}
\usepackage{longtable}
\usepackage{float}
\usepackage{url}
\usepackage{hyperref}
\usepackage{amsthm}
\usepackage{float}
\usepackage{xcolor}
\usepackage{comment}

\newcommand{\be}{\begin{equation}}
\newcommand{\ee}{\end{equation}}
\newcommand{\bea}{\begin{eqnarray}}
\newcommand{\eea}{\end{eqnarray}}

\begin{document}

\title{Cosmic Collider Gravitational Waves sourced by \\ Right-handed Neutrino production from Bubbles: \\ \it{Testing Scales of Seesaw, Leptogenesis and Dark Matter}}

\author{Anish Ghoshal}
\affiliation{Department of Physics and Astronomy, University of Sussex,
Brighton, BN1 9RH, United Kingdom}

\author{Pratyay Pal}
\email{anish.ghoshal@fuw.edu.pl}
\email{pratyaypal2025@gmail.com}
\affiliation{Department of Physical Sciences, Indian Institute of Science Education and Research Kolkata, Mohanpur-741 246, WB, India}

\begin{abstract}
We study a minimal type-I seesaw framework in which a first-order phase transition (FOPT), driven by a singlet scalar $\phi$, produces right-handed neutrinos (RHNs) through bubble collisions, realizing a \emph{cosmic-scale collider} that probes ultra-high energy scales. The resulting inhomogeneous RHN distribution sources a novel low-frequency gravitational-wave (GW) signal in addition to the standard bubble-collision contribution. A stable lightest RHN can account for the observed dark matter (DM) relic abundance for masses as low as $M_{1} \equiv m_{\rm DM} \gtrsim 10^{6}\,\mathrm{GeV}$, with the associated novel GW signal accessible in LISA, ET and upcoming LVK detectors. If the RHNs are unstable, their CP-violating decays generate the baryon asymmetry via leptogenesis for $M_{1} \gtrsim 10^{11}\,\mathrm{GeV}$ and phase transition temperatures $T_* \gtrsim 10^{6}\,\mathrm{GeV}$, yielding the novel GW signatures within the reach of ET, BBO and upcoming LVK detectors. Part of the parameter space is already constrained by the LVK $\mathcal{O}(3)$ data. If RHN decays populate a dark sector fermion $\chi$ with mass $m_{\chi} \in [10^{-4},10^{4}],\mathrm{GeV}$, successful co-genesis of baryons and asymmetric dark matter is achieved for $T_* \gtrsim 10^{7}\,\mathrm{GeV}$ and $M_{1} \gtrsim 10^{9}\,\mathrm{GeV}$, naturally explaining $\Omega_{\rm DM} \simeq 5\Omega_{\rm B}$. The corresponding GW signals are testable with LISA, ET, and BBO. Finally, we analyze a UV-complete multi-Majoron model based on a global $U(1)_N \times U(1)_{\rm B-L}$ extension of Standard Model (SM), motivated from the hierarchy of lepton masses, in which a distinctive GW signature associated with a \emph{cosmic Majoron collider} arises from scalar production during $U(1)_N$ symmetry breaking, detectable by BBO, ET and  upcoming LVK. Successful leptogenesis is realized for the heaviest RHN mass $M_3 \sim 10^{10}\,\mathrm{GeV}$ and a $U(1)_N$ breaking vev $v_2 \sim \mathcal{O}(\mathrm{TeV})$ that sets the seesaw scale.
\end{abstract}

\maketitle

\tableofcontents

\section{Introduction}

Cosmological first-order phase transitions (FOPTs) involve the nucleation of true-vacuum bubbles in a metastable false-vacuum background~\cite{Hogan:1983ixn, Witten:1984rs, Hogan:1986qda, Kosowsky:1991ua, Kamionkowski:1993fg, Kosowsky:1992vn}. When these bubbles expand and collide, they generate a large stochastic gravitational-wave (GW) background that can be tested in upcoming GW missions~\cite{Grojean:2006bp,Caprini:2015zlo,Caprini:2018mtu,Caprini:2019egz,Athron:2023xlk}.

 In certain classes of FOPTs, the bubble walls may accelerate to ultra-relativistic speeds, the so-called runaway scenario. This typically occurs during supercooled transitions, transitions into cold sectors, or sectors lacking gauge bosons. In such cases, plasma friction becomes negligible due to plasma dilution, and a large fraction of the released vacuum energy accumulates in the bubble walls. Consequently, the GW signal is dominated by the scalar field energy in the walls after collision~\cite{Kosowsky:1991ua,Kosowsky:1992rz,Kosowsky:1992vn,Kamionkowski:1993fg,Caprini:2007xq,Huber:2008hg,Bodeker:2009qy,Jinno:2016vai,Jinno:2017fby,Konstandin:2017sat,Cutting:2018tjt,Cutting:2020nla,Lewicki:2020azd}.

Beyond the classical picture of bubble dynamics and energy dissipation~\cite{Hawking:1982ga,Kleban:2011pg,Chang:2008gj}, ultra-relativistic walls can also produce particles with energies far above the ambient plasma temperature, an inherently quantum effect analogous to particle production from vacuum. Collisions of such vacuum bubbles in the early Universe thus serve as as cosmic-scale high-energy colliders with energy even upto the Planck scale. These
“cosmic colliders”\footnote{Not to be confused with ``Cosmological Collider" which involves particle production during inflation from inflationary fluctuations \cite{Arkani-Hamed:2015bza}.} are then possibly the most energetic phenomena in our universe reached ever \cite{Shakya:2025qpi}, and have energies larger than any energy scale reachable in traditional laboratory colliders. On microscopic scales (set by the Lorentz-contracted bubble-wall thickness), bubble collisions can efficiently convert vacuum energy into energetic particles~\cite{Jinno:2017fby,Konstandin:2011ds,Falkowski:2012fb,Watkins:1991zt}. Recent work~\cite{Shakya:2023kjf,Mansour:2023fwj} shows that this production follows a universal power law that is largely insensitive to the details of the collision, provided the walls are ultra-relativistic. If the produced particles survive long enough to generate a quadrupole anisotropy, they can source an additional GW component~\cite{Inomata:2024rkt}. Such GWs can probe fundamental particle-physics properties and offer new windows into early-Universe dynamics.

Observations of neutrino oscillations~\cite{Fukuda_2001, Fukuda_2002, Ahmad_2002, Ashie_2005, Abe_2016, Smirnov_2016, Aartsen_2018, Abe_2008, Abe_2011} imply that neutrinos possess non-zero masses, requiring physics beyond the Standard Model (SM). Introducing two or more right-handed Majorana neutrinos (RHNs) leads to the Type-I seesaw mechanism~\cite{Minkowski:1977sc,Yanagida:1979as,Glashow:1979nm,Mohapatra:1979ia}, which naturally explains small neutrino masses. The same RHNs can also account for the observed baryon asymmetry of the Universe (BAU)~\cite{Planck2018,ParticleDataGroup:2020ssz} through the leptogenesis mechanism~\cite{Fukugita:1986hr}. In conventional thermal leptogenesis, successful asymmetry generation requires the lightest RHN to satisfy the Davidson–Ibarra bound, $M_1 \gtrsim 10^9$ GeV, assuming the RHN mass spectrum is hierarchical, along with no lepton flavor effects~\cite{Nardi:2006fx,Abada:2006fw,Abada:2006ea,Giudice:2003jh}. Such high scales challenge direct tests in laboratory experiments, however, certain indirect signatures of new physics like lepton number violation processes through neutrinoless double beta decay~\cite{Cirigliano:2022oqy} or CP violation in neutrino oscillation~\cite{Endoh:2002wm} can be looked for. Moreover, theoretical constraints on low-energy couplings can arise from demanding consistency with UV-complete frameworks, such as $SO(10)$ Grand Unified Theories (GUTs)~\cite{DiBari:2008mp,Bertuzzo:2010et,Buccella:2012kc,Altarelli:2013aqa,Fong:2014gea,Mummidi:2021anm,Patel:2022xxu}. There may be additional consistencies related to the demand of electroweak (EW) SM Higgs vacuum (meta)stability in the early Universe~\cite{Ipek:2018sai,Croon:2019dfw}. All of these provide some useful bounds on the large parameter space involved in Seesaw and Leptogenesis scenarios. To further make these bounds sharper and probe the relevant high scales of leptogenesis, we allude to non traditional or non laboratory based searches, particularly cosmological observables as tools to see indirect signatures or tests for such scales of new physics involving seesaw, leptogenesis, and heavy dark matter etc. 

The other major puzzle is the nature and origin of dark matter (DM), which constitutes about $\sim 27\%$ of the Universe’s energy density~\cite{Zwicky:1933gu, Zwicky:1937zza, Rubin:1970zza, Clowe:2006eq}. In terms of density 
parameter $\Omega_{\rm DM}$ and reduced Hubble constant $h = \text{Hubble Parameter}/(100 \;\text{km} ~\text{s}^{-1} 
\text{Mpc}^{-1})$, the current DM abundance is usually depicted to be \cite{Aghanim:2018eyx}

\begin{equation}
\Omega_{\text{DM}} h^2 = 0.120\pm 0.001
\label{dm_relic}
\end{equation}

\noindent at 68\% CL. Meanwhile the visible or baryonic matter content constitutes about $\sim 5\%$ of the Universe’s energy density, which is asymmetric in nature - there is huge excess of baryon (matter) than anti-baryons (antimatter). 

\medskip

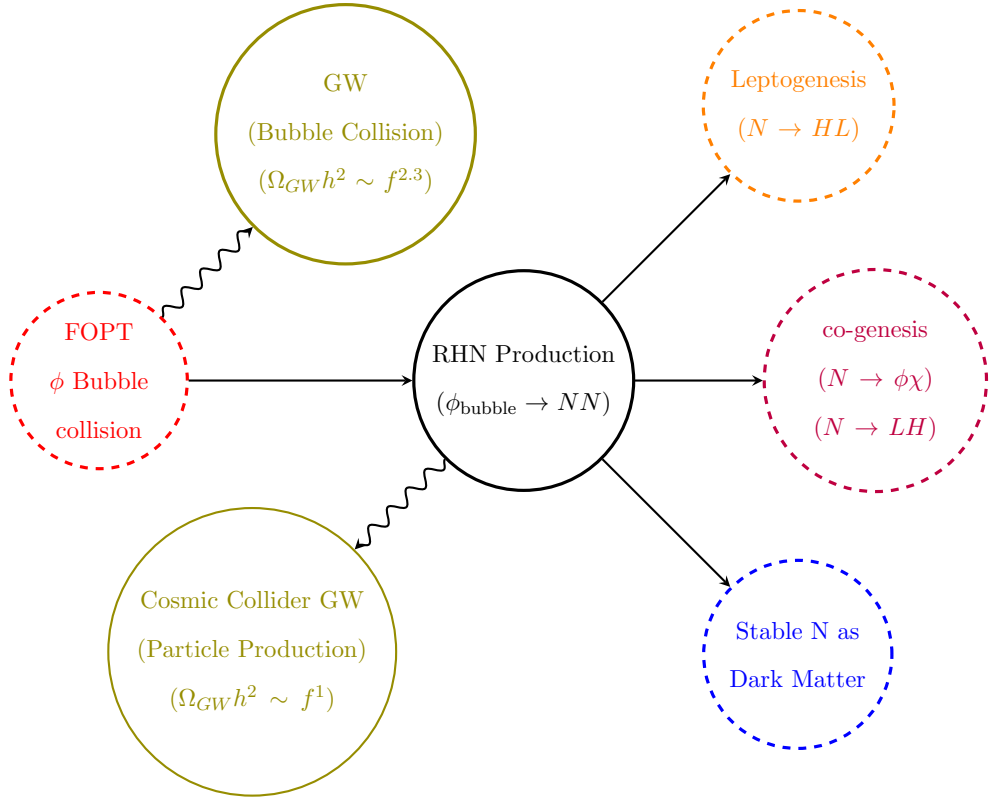
\begin{figure}[H]
\centering
\begin{tikzpicture}[scale=0.85, every node/.style={transform shape}, >=stealth, node distance=2.8cm]

\node[
    draw, circle, dashed, red, very thick,
    minimum size=1.9cm,
    text width=1.6cm,
    align=center
] (I) {FOPT\\ $\phi$ Bubble collision};

\node[
    draw, circle, above right = 2cm of I,
    olive, very thick,
    minimum size=1.9cm,
    text width=3.2cm,
    align=center
] (BC) {GW\\ (Bubble Collision)\\ $(\Omega_{GW}h^2\sim f^{2.3})$};

\node[
    draw, circle,
    right=3.5cm of I,
    very thick,
    minimum size=2.3cm,
    text width=2.9cm,
    align=center
] (R) {RHN Production\\$(\phi_{\rm bubble}\rightarrow NN)$ };

\node[
    draw, circle, below left= 2cm of R,
    olive, thick,
    minimum size=1.9cm,
    text width=3.7cm,
    align=center
] (PP) {Cosmic Collider GW\\ (Particle Production)\\ $(\Omega_{GW}h^2\sim f^1)$};

\node[
    draw, circle,
    above right=2cm and 2cm of R,
    dashed, orange, very thick,
    minimum size=1.7cm,
    text width=2.4cm,
    align=center
] (SM) {Leptogenesis\\$(N\rightarrow HL)$};

\node[
    draw, circle,
    below right=2cm and 2cm of R,
    dashed, blue, very thick,
    minimum size=1.7cm,
    text width=2.4cm,
    align=center
] (DM) {Stable N as Dark Matter};

\node[
    draw, circle,
    right=2cm of R,
    dashed, purple, very thick,
    minimum size=1.7cm,
    text width=2.5cm,
    align=center
] (CG) {co-genesis\\$(N\rightarrow \phi\chi)$\\$(N\rightarrow LH)$};

\draw[->, thick] (I) -- (R);

\draw[->, thick, decorate, decoration={snake}] (I) -- (BC);

\draw[->, thick, decorate, decoration={snake}] (R) -- (PP);

\draw[->, thick] (R) -- (SM);

\draw[->, thick] (R) -- (DM);

\draw[->, thick] (R) -- (CG);

\end{tikzpicture}
\caption{\it 
Schematic diagram for Cosmic Collider~\cite{Shakya:2025qpi} during runaway FOPTs showing the correlation between RHN production, Dark matter and Leptogenesis along with the resulting cosmological signals - the well established bubble collision GWs as well as the novel RHN production GWs as discussed in this analysis. 
}
\label{fig:diag}
\end{figure}

The observed BAU is quantitatively depicted as the ratio of excess of baryons over anti-baryons to photon \cite{Aghanim:2018eyx} 
\begin{equation}
\eta_B = \frac{n_{B}-n_{\overline{B}}}{n_{\gamma}} \simeq 6.2 \times 10^{-10}
\label{etaBobs}
\end{equation}
as observed from the cosmic microwave background (CMB) measurements, and from the big bang nucleosynthesis (BBN) estimates ~\cite{Zyla:2020zbs}. The numerical proximity of DM and baryon abundances $\Omega_{\rm DM} \approx 5\,\Omega_{\rm Baryon}$, motivates scenarios where both originate from a common mechanism (see Ref.~\cite{Boucenna:2013wba} for a review on this topic). This has led to many proposals where the usual mechanism for baryogenesis directly or via leptogenesis is extended to the dark sector assuming the dark sector to be asymmetric \cite{Nussinov:1985xr, Davoudiasl:2012uw, Petraki:2013wwa, Zurek:2013wia}. In typical asymmetric dark matter (AsDM) scenario, the same out-of-equilibrium decay of a heavy particle, for instance, the decay of heavy right-handed neutrinos (RHN) into SM leptons and dark sector leads to the creation of asymmetries in the two sectors of similar order of magnitudes $n_B-n_{\overline{B}} \sim \lvert n_{\rm DM}-n_{\overline{ \rm DM}} \rvert$ \cite{Falkowski_2011,Barman_2022}.
\medskip

 In order to understand the impact of the presence of such heavy neutrinos in the theory, we investigate FOPT in a minimal scalar singlet ($\phi$) extended SM. During the phase transition, the bubble walls of $\phi$ decays to RHNs. The subsequent decays of these right-handed neutrinos are responsible for baryogenesis via leptogenesis \cite{Fukugita:1986hr}. We also study a simplistic possibility where one of the right-handed neutrino is stable and be the DM candidate. Finally we also investigate the scenario where RHN decays can also furnish a dark matter production mechanism \cite{Falkowski_2011, Barman_2022}, which can result in asymmetric dark matter.
 \vspace{0.4cm}

\medskip

 Along with this we offer a possibility of understanding the GW production that occur during the FOPT, particularly from RHN production (besides the standard GW from bubble wall production) as a novel probe of non-thermal leptogenesis and DM formation. The signature in the GW spectrum follows a characteristic power law for RHN production and in some parameter space can give rise to a louder signal than the bubble collision one, which may enable us to detect in future the GW from RHN production in some of the upcoming GW missions and its impact on understanding the DM and matter-antimatter asymmetry puzzles. A schematic diagram for the overview of our framework is given in Fig.~\ref{fig:diag}.

\medskip

The paper is organised as follows: In sec~\ref{II} we discuss the particle production formalism from an early FOPT. Sec~\ref{III} describes the Gravitational Waves produced during a FOPT from different sources and their detection prospects. In sec~\ref{IV}, we discuss the possibility of producing various cosmological relics such as DM and Baryon Asymmetry from the production of RHNs during bubble wall decay. We also show possible GW detection scenarios for both DM and leptogenesis. In sec.~\ref{V}, we examine the co-genesis framework in which both DM and the baryon asymmetry are generated simultaneously via RHN decays, together with the resulting gravitational-wave signatures. In sec.~\ref{VI}, we calculate the phase transition parameters coming from a FOPT in an UV complete multi-Majoron model, showing different parameter space where the distinct GW signal coming from particle production can trace non-thermal leptogenesis, providing us a proof of the concepts presented in the previous sections.  Finally in sec~\ref{conclusion} we summarize our results.

\medskip

\section{Particle Production from Bubbles during first-order Phase Transition}
\label{II}

During a FOPT bubble wall collisions can lead to production of various particles coupled to the scalar field . Depending on the details of the scalar potential, the collisions can be elastic or inelastic. In both cases, the scalar field at the point of collision gets excited to a field value away from the minima. This results in oscillations around true (false) minimum for inelastic (elastic) collisions. Such dynamics of the background field then can give rise to significant particle production to any field that couples to it leading to a significant fraction of the vacuum energy getting converted into particle population \cite{cataldi2024leptogenesisbubblecollisions,shakya2025aspectsparticleproductionbubble,Watkins:1991zt}.

\subsection{Formalism}
\noindent The formalism consists of treating the moving and colliding bubble walls and subsequent oscillations as classical external field configurations of the scalar $\phi(x,t)$.  The probability for this configuration to decay
into particles is extracted from the imaginary part of its effective action \cite{Watkins:1991zt}, 
\begin{equation}
    P_{\geq1} = 2\ \text{Im}(\Gamma[\phi])
\end{equation}
where, $P_{\geq 1}$ denotes the probability of producing atleast one particle after collision and the effective action is written as 
\begin{equation}
    \Gamma[\phi] = \sum_{n = 2}^\infty \frac{1}{n!}\int d^4 x_1...d^4x_n \ \Gamma^{(n)}(x_1,...,x_n) \phi(x_1)...\phi(x_n) .
\end{equation}
Here, $\Gamma^{(n)}$ is the n-point 1-PI effective action. Assuming $P_{\geq1} \ll 1$, to the leading order we find,
\begin{equation}
    \Gamma[\phi] = \frac{1}{2}\int d^4x_1d^4x_2\ \phi(x_1)\phi(x_2)\Gamma^{(2)}(x_1,x_2).
\end{equation}
Hence, 
\begin{equation}
    \text{Im}(\Gamma[\phi]) = \frac{1}{2} \int d^4x_1d^4x_2\phi(x_1)\phi(x_2) \int \frac{d^4p}{(2\pi)^4} e^{ip\cdot(x_1-x_2)} \text{Im}\left(\tilde{\Gamma}^{(2)}(p^2)\right)
    \label{4}
\end{equation}
where $\tilde{\Gamma}^{(2)}$ is the Fourier transform of the 2 point 1-PI effective action $\Gamma^{(2)}(x_1,x_2)$. Taking $\phi(x_1) = \int \frac{d^4k_1}{(2\pi)^4}e^{-ik_1\cdot x_1}\tilde{\phi}(k_1)$ and inserting into Eq.~(\ref{4}) we find 
\begin{equation}
    \text{Im}(\Gamma[\phi]) = \int\frac{d^4 p}{(2\pi)^4} |\tilde{\phi}(p)|^2 \text{Im}\left(\tilde{\Gamma}^{(2)}(p^2)\right).
    \label{5}
\end{equation}
This formula has a simple interpretation. The Fourier transform decomposes the scalar field into modes of definite four-momentum $p^{\mu}$. Modes with $p^2 > 0$ represents propagating “particles” with mass $M^2 = p^2$, which in general can be off-shell. Eq.(\ref{5}) sums over the number of $\phi$ particles with mass M contained in the field multiplied by the probability for those particles to decay. \\

\noindent In the case of plane-symmetric walls moving in the z-direction, one can write $\phi(x,t) = \phi(z,t)$. The corresponding Fourier transform takes the form, $\tilde{\phi}(\vec{k},\omega) = (2\pi)^2\delta(k_x)\delta(k_y) \tilde{\phi}(k_z,\omega)$. Substituting this in Eq.~(\ref{5}) and integrating over $p$ we get the number of particles produced per unit area $N/A$, given by -
\begin{equation}
    \frac{N}{A} = 2\int \frac{dk d\omega}{(2\pi)^2} |\tilde{\phi}(k,\omega)|^2 \text{Im}\left[\tilde{\Gamma}^{(2)}(\omega^2-k^2)\right].
\end{equation}
\noindent Here modes of off shell $\phi$ quanta propagating with $m^2 = \omega^2-k^2 = \chi$ are decaying into other particles. To obtain the number density of these produced particles, first we note that, from the Optical Theorem, the imaginary part of the 2-point 1PI Green's function is given by the sum over matrix elements of all possible decay processes:
\begin{equation}
    \text{Im}\left[\tilde{\Gamma}^{(2)}(\chi)\right] = \frac{1}{2}\sum_\alpha\int d\Pi_\alpha |\mathcal{M}(\phi\rightarrow\alpha)|^2\Theta(\chi - \chi_{min(\alpha)})
    \label{7},
\end{equation}
where $\Theta$ is the heaviside step function and the sum runs over all possible final states $\alpha$ that can be produced. $|\mathcal{M}(\phi\rightarrow\alpha)|^2$ is the spin averaged squared amplitude and $\chi_{min(\alpha)} = (\sum_\alpha m_\alpha)^2$ represents the minimum energy
required to produce the final state particles on-shell. 
For n-body final states, one should replace the prefactor 2
in Eq.~(\ref{7}) by the appropriate number $1/n!$. Going to $\chi = \omega^2 - k^2$ and $\xi = \omega^2+k^2$ variables and integrating Eq.~(\ref{7}) we get :
\begin{equation}
    \frac{N}{A} = \frac{1}{2\pi^2}\int_{p_{min}^2}^{p_{max}^2} dp^2 f(p^2)\text{Im}\left[\tilde{\Gamma}^{(2)}(p^2)\right]\ ; \quad f(\chi) = \int_\chi^{2\chi_{max}-\chi}d\xi \frac{1}{\sqrt{\xi^2-\chi^2}}|\tilde{\phi}(\xi,\chi)|^2.
    \label{8}
\end{equation}
Here $f(p^2)$ is known as the efficiency factor and encapsulates the relevant details of the underlying field configuration. The lower limit of the integral is determined by either the mass of the particle species being produced - $p_{min} = 2m$ for pair production, or by the inverse size of the bubble - $p_{min} = 2R_*^{-1}$ where $R_*$ is the colliding bubble radius. The upper cutoff is provided by $p_{max} = 2/l_w$, the energy in the two colliding bubble walls, which represents the maximum energy available in the process. $l_w$ here is the Lorentz contracted wall thickness. Similarly one can get the expression for the energy density of the produced particles per unit area $(E/A)$ as 
\begin{equation}
\frac{E}{A} = \frac{1}{2\pi^2} \int_{p_{min}^2}^{p_{max}^2} dp^2\ p f(p^2)\text{Im}\left[\tilde{\Gamma}^{(2)}(p^2)\right] .
\label{9}
\end{equation}
The produced number density can also be calculated, 
\begin{equation}
    n = \frac{N}{A}\times\frac{2\pi R_*^2}{\frac{4}{3}\pi R_*^3} = \frac{3}{4\pi^2 R_*}\int_{p_{min}^2}^{p_{max}^2} dp^2 f(p^2)\text{Im}\left[\tilde{\Gamma}^{(2)}(p^2)\right].
    \label{10}
\end{equation}
Note in one collision, per bubble, only half of its surface area effectively collides.
\subsection{The Efficiency Factor}
\noindent The efficiency factor $f(p^2)$  depends on the nature of the collision and the subsequent dynamics of the background field and has been calculated in the literature. We quote the results here for a runaway phase transition with negligible plasma friction, which will be used throughout this work \cite{Mansour:2023fwj}:
\begin{align}
    f_{elastic}(p^2) &= f_{PE}(p^2) + \frac{v_\phi^2L_p^2}{15m_t^2}\exp\left(\frac{-(p^2-m_t^2+12 m_t/L_p)^2}{440 m_t^2/L_p^2}\right).\\[0.2cm]
    f_{inelastic}(p^2) &= f_{PE}(p^2) + \frac{v_\phi^2L_p^2}{4m_f^2}\exp\left(\frac{-(p^2-m_f^2+ 31 m_f/L_p)^2}{650 m_f^2/L_p^2}\right).
\end{align}
Here $m_t$, $m_f$ are the scalar masses in the true and false vacua respectively. $L_p = \min(R_*,\Gamma_{\phi}^{-1})$ where $\Gamma_\phi$ is the total decay rate of the scalar as it performs oscillations around its true or false minimum. Note $\langle\phi\rangle = v_\phi$ is the vev of the scalar field, $\gamma_w$ is the boost factor of the bubble wall at collision and $l_w = l_{w0}/\gamma_w$ is the Lorentz contracted wall width at the collision. Finally, $f_{PE}$ is the efficiency factor for a perfectly elastic collision and is given as
\begin{equation}
    f_{PE}(p^2) = \frac{16v_\phi^2}{p^4}\log\left[\frac{2(\gamma_w/l_{w0})^2-p^2+2(\gamma_w/l_{w0})\sqrt{(\gamma_w/l_{w0})^2-p^2}}{p^2}\right].
    \label{12}
\end{equation}
We see this has an approximately power law component $f_{PE} \sim p^{-4}$, which originates from the nontrivial dynamics of the background field when the bubbles collide. The remaining part in $f(p^2)$ behaves approximately as a Gaussian peak centered around the mass of the scalar in the relevant vacuum.  This comes from the oscillation of the scalar field around its relevant minimum after the collision. We note~\cite{shakya2025aspectsparticleproductionbubble} from numerical estimates, the log factor in Eq.~(\ref{12}) ranges between 6 $\sim$ 60 for a wide range of phase transition parameter values and hence can be treated as a constant.

\subsection{Effective Action Calculation}

\noindent In the formalism above, in Eq.~(\ref{8},\ref{9}), the particle physics information is encoded in the 2-point 1PI Green's function $\tilde{\Gamma}^{(2)}$, to which we now turn our attention. Following Eq.~(\ref{7}) we note, to calculate the overall decay probability of the background field, we need to calculate $|\mathcal{M}(\phi\rightarrow\alpha)|^2$ for all particle combinations $\alpha$ that are allowed in the setup. However, to calculate the decay probability into a given final state important for our purpose, it is sufficient to perform the calculation solely for this channel. The full sum over all final states is not required as long as the total decay probability remains less than unity. This ensures that no individual decay channel induces a significant backreaction on the system. With these considerations in mind, we calculate Eq.~(\ref{7}) for scenarios where fermions are produced from bubble collisions :
\begin{enumerate}    
    \item \textbf{Fermions}: Fermions can be produced via some Yukawa coupling to the scalar $\phi$ of the form $\mathcal{L}_I \supset -y_f\phi\bar{\psi}\psi$. This gives rise to $\phi^*\rightarrow \bar{\psi}\psi$ decay process~\cite{Mansour:2023fwj}:
    \begin{equation}
        \text{Im}\left[\tilde{\Gamma}^{(2)}(p^2)\right]_{\phi^*\rightarrow\bar{\psi}\psi} = \frac{y_f^2 p^2}{8\pi}\left(1-\frac{4m_\psi^2}{p^2}\right)\Theta(p-2m_\psi).
    \end{equation}
    Note here the mass $m_\psi$ refers to the mass of the particle in true vacuum which after the scalar acquires a vev $v_\phi$ becomes $m_\psi = m_{\psi,0} + y_f v_\phi$.
\end{enumerate}

\noindent We note here, that scalar and gauge field particles can also be produced from the bubble wall collisions, see Appendix~\ref{app-A}. However, in the remainder of this work starting from chapters~\ref{IV}\& \ref{V}, we will primarily focus on fermion production since we are interested in heavy Right Handed Neutrino production.

\medskip

\section{Gravitational Waves from FOPT during Particle Production}
\label{III}
\noindent Gravitational waves (GWs) from first-order phase transitions (FOPTs) in the early Universe \cite{Witten:1984rs,Hogan:1986qda,PhysRevLett.65.3080,Kamionkowski_1994} have been widely
studied as well-motivated and promising targets for current and upcoming GW detectors \cite{amaroseoane2017laserinterferometerspaceantenna,Kawamura:2006up,Punturo:2010zz}. Although the
electroweak phase transition and the QCD phase transition within the Standard Model (SM) are known to not be
first-order\footnote{Though some new possibilities has been recently proposed in literature, see Ref.~\cite{Shakya:2025mdh,Yamada:2025hfs,Yamada:2025cfr}.}, the detection of a cosmological FOPT would be a smoking gun for physics beyond the Standard Model
(BSM). Indeed, there are numerous BSM frameworks that predict FOPTs in dark sectors \cite{Schwaller_2015,Caprini_2018,Baldes_2017,Hall_2023,Croon_2018,Prokopec_2019} which can produce
observable GW signals. For a detailed summary of the GW formalism, we refer the reader to Appendix~\ref{app-A} and proceed with GW spectrum calculation next.

\subsection{Relevant Phase Transition Parameters}
\noindent Here we list all the key parameters relevant for the production of GWs and particles during a FOPT:  
\begin{itemize}
    \item \textbf{$T_n$}: Temperature of the thermal bath at which the FOPT is triggered, i.e., when bubbles of the true vacuum begin to nucleate at a rate exceeding the Hubble expansion rate.
    
    \item \textbf{$R_0$}: Critical radius of a nucleated bubble that can successfully grow. Typically, $R_0 \sim \mathcal{O}(T_n^{-1})$.
    
    \item \textbf{$\alpha$}: Strength of the phase transition, defined as 
    \[
    \alpha \equiv \frac{\rho_{\text{vac}}}{\rho_{\text{rad}}} = \frac{\Delta V}{\rho_{\text{rad}}},
    \]
    where $\rho_{\text{vac}} = \Delta V$ is the vacuum energy difference and $\rho_{\text{rad}}$ is the radiation energy density of the thermal bath at $T_n$.
    
    \item \textbf{$\beta$}: (Inverse) duration of the phase transition. It is commonly expressed in units of the Hubble rate $H$ as the dimensionless ratio $\beta/H$.
    
    \item \textbf{$v_w$}: Velocity of the bubble wall. This quantity evolves with time: as the bubble expands, vacuum energy is transferred to the wall, accelerating it. The velocity $v_w$ can approach a terminal value if frictional forces from the plasma become significant.
    
    \item \textbf{$\gamma_w$}: Lorentz boost factor of the bubble wall, related to its velocity by $\gamma_w = 1/\sqrt{1 - v_w^2}$.
    
    \item \textbf{$l_w$}: Thickness of the bubble wall. Initially, at nucleation, $l_{w0} \sim \mathcal{O}(v_\phi^{-1})$. As the wall accelerates, the apparent thickness in the plasma frame is Lorentz contracted, $l_w = l_{w0}/\gamma_w$, and thus tends to decrease with time.
    
    \item \textbf{$R_*$}: Typical bubble size at collision. This is determined by the duration of the transition and is approximately given by
    \[
    R_* \simeq (8\pi)^{1/3} \, \frac{v_w}{\beta}.
    \]
    
    \item \textbf{$T_*$}: Temperature of the thermal bath at which the bubbles percolate and the phase transition completes. For a radiation-dominated Universe, $T_* \simeq T_n$. If the Universe becomes vacuum dominated, $T_*$ is determined through energy conservation at the end of the transition.
\end{itemize}
We will be using these parameters throughout the rest of this paper, in expression of GW spectrum as well as in the calculation of various cosmological relic abundances.

\subsection{GW From Bubble Collision}
\noindent FOPTs proceed via the nucleation of bubbles,
their expansion, collision and thermalization into light
particles, and GWs are produced during this process~\cite{Caprini_2016}. In
the transition process, some of the released energy goes into
heating up the plasma, while the rest is mostly carried by the scalar
field configuration (bubble wall) and/or the bulk motion of
the surrounding fluid. Gravitational wave production by
such localized structure of energy around the walls are studied in detail over the past decade in literature and recent developments can be found in Refs.~\cite{Jinno_2017,Jinno_2019,Cutting_2018}. For runaway bubble configurations, GWs are primarily sourced by bubble wall collisions, i.e. the scalar field and produced particle densities. The GW spectrum obtained from the study in \cite{Caprini_2024,ghoshal2025complementaryprobeswarpedextra} is given as 
\begin{equation}
    \Omega_{GW}h^2(f) \simeq 16\frac{(f/f_p)^{2.4}}{[1+(f/f_p)^{1.2}]^4}\bar{\Omega}_{GW}h^2,
    \label{27}
\end{equation}
where $h \equiv 10^{-2}H_0\ Mpc/(km/s) \simeq 0.674$ is the dimensionless parameter of the Hubble constant today, $H_0$, while $\bar{\Omega}_{GW}h^2$ and $f_p$ are respectively the spectrum peak’s amplitude and frequency given by
\begin{align}
    \bar{\Omega}_{GW}h^2 &\simeq 3.8\times10^{-6}\left(\frac{\beta}{H}\right)^{-2}\left(\frac{\alpha}{1+\alpha}\right)^{2}\frac{1}{g_*^{1/3}(T_*)},\\
    f_p &\simeq 8.4\times 10^{-7}\left(\frac{\beta}{H}\right)g_*^{1/6}(T_*)\left(\frac{T_*}{100\ \text{GeV}}\right).
\end{align}
Here $g_*$ is the total number of degrees of freedom present in the plasma. 
The SGWB spectrum then exhibits a broken-power-law frequency shape. In particular\footnote{This is much steeper than what Ref~\cite{Caprini_2020} considered, i.e. $\Omega_{GW}h^2\propto f^{-1}$.}\footnote{With an expanding FRW background with more recent developments in lattice simulation, the scaling has been found to be slightly different in Ref.~\cite{Lewicki:2025hxg}.}, at $f\ll f_p$ it behaves as $\Omega_{GW}h^2\propto f^{2.3}$  and at $f\gg f_p$, it behaves as $\Omega_{GW}h^2\propto f^{-2.4}$.  

We will see in sec-\ref{IV} and sec-\ref{V} that for most of our analysis we will be in this regime only and hence we will be using Eq.~(\ref{27}) for calculating the bubble wall contribution to the GW spectrum observed today.

\subsection{GW From Particle Production}

\noindent In sec-\ref{II} we saw that bubbles produced during a FOPT can create a particle density when they collide. If the energy momentum tensor of these produced particles then have a transverse traceless component, they can also source GWs. Indeed this contribution can be estimated via calculating the UETC in Eq.~(\ref{24}) for the produced particle distribution. This would be different from the UETC of the Bubble collisions. Then substituting the UETC in Eq.~(\ref{26}) and performing the integral one can get the GW spectrum coming from particles. For the particle production mechanism, this contribution is equal to~\cite{Inomata:2024rkt}
\begin{equation}
    \Omega_{GW}h^2(f) = 1.65\times 10^{-5} \left(\frac{\beta}{H}\right)^{-2}\left(\frac{\alpha}{1+\alpha}\right)^{2}\left(\frac{g_*(T_*)}{100}\right)^{1/3}\Delta_{GW}^{pp}(f),
    \label{30}
\end{equation}
where $\Delta_{GW}^{pp}$ has the information of the UETC and is numerically estimated to be:
\begin{equation}
    \Delta_{GW}^{pp\ (fit)}(k/\beta) \approx \frac{0.003\kappa^2 k/\beta}{1+ 16(k/\beta)^3} \qquad (\text{for}\ k/\beta\lesssim 0.75).
    \label{36}
\end{equation}
Here $k$ is the momentum of the mode and $\kappa$ is the efficiency factor -\\
\begin{equation}
    \kappa \equiv \frac{\frac{1}{2}(E/A)\cdot4\pi R_*^2}{\frac{4}{3}\pi R_*^3\cdot\Delta V} = \frac{3}{2}\frac{E/A}{R_*\Delta V}.
    \label{38}
\end{equation}

\noindent The momentum $k$ is related to the observed GW frequency $f$ today via the relation 

\begin{equation}
    f = \frac{k}{2\pi}\left(\frac{a_*}{a}\right) = 2.63\times10^{-6}\ \text{Hz}\times\left(\frac{k}{\beta}\right)\left(\frac{\beta}{H}\right)\left(\frac{T_*}{100\ GeV}\right)\left(\frac{g_*(T_*)}{100}\right)^{1/6}.
\end{equation} 

\vspace{0.3cm}

\noindent Note in Eq.~(\ref{38}), the numerator represents the amount of energy that goes into the produced particles and the denominator is the total vacuum energy released as the bubble expanded to its collision size. Here we have assumed the released vacuum energy gets completely transferred into the bubbles which subsequently collides and gives $\kappa$ fraction of its energy to particle production. 
The factor of 1/2 in the numerator accounts for the fact that particle production at collision comes from the energy released from two bubbles.\\

\medskip

\noindent The particular value of $\kappa$ would depend on the microphysics of the interaction, namely the couplings of the Lagrangian and also on the phase transition parameters. We need to specify the interaction to be able to calculate it. But as will be shown in sec-\ref{IV}, for our physics scenarios we can safely take $\kappa\sim\mathcal{O}(1)$.\\

\medskip

\begin{figure}[H]
    \centering
    \includegraphics[width=0.75\linewidth]{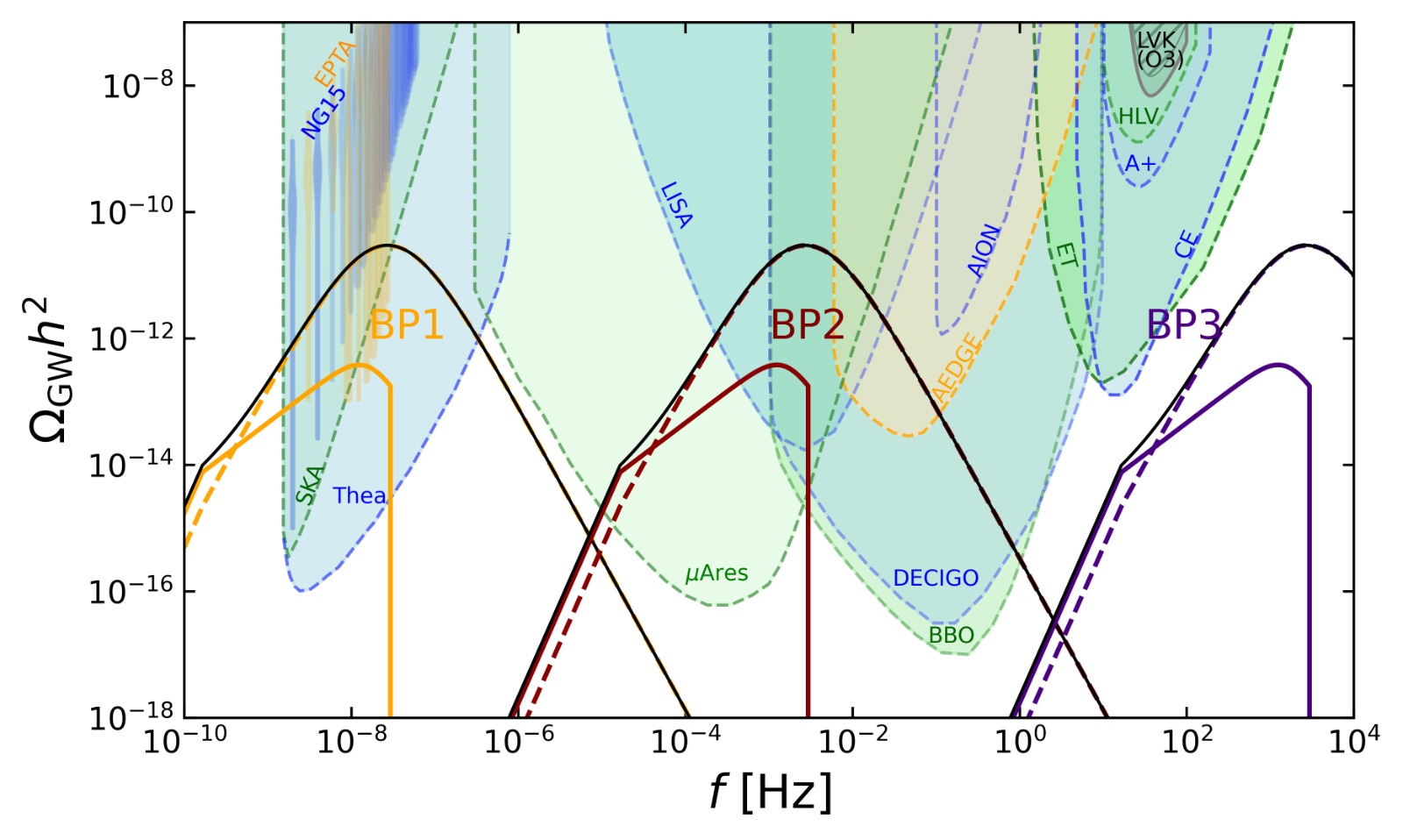}
    \caption{\it GW spectrum plotted against power-law integrated sensitivity curves for various present and upcoming GW detectors for $\alpha = 10$, $\beta/H = 150$. Colored solid lines represent the GW spectrum from particle production, while dashed lines correspond to the contribution from bubble collisions. The black solid lines denote the total GW spectrum obtained by combining both sources. The three sets of curves correspond to phase transitions at temperatures - \textbf{BP1} ($T_* = 10^{-5}$ TeV), \textbf{BP2} ($T_* = 1$ TeV) and \textbf{BP3} ($T_* =10^6$ TeV) (orange, red, indigo), respectively. In calculating particle production contribution we have taken $\kappa \sim 1$ as discussed. }
    \label{fig:1}
\end{figure}

\noindent Eq.~(\ref{36}) is robust for $k\lesssim \beta$ and runs into problematic behavior for $k\gtrsim \beta$. This originates from the short-distance structure of the two-point correlation function of the source in configuration space: as two bubbles cannot nucleate at the same spatial point, the correlation function must vanish in the limit of zero separation. This leads to a negative power spectrum in Fourier space for some frequencies in large momenta regime. We consequently discard this unphysical region and focus on the large scale physics for $k\lesssim \beta$.  The expression Eq.~(\ref{36}) remains reliable up to $k/\beta \lesssim 0.75$ \cite{Inomata:2024rkt}. In our analysis, we therefore present the GW spectrum only within this domain and truncate it beyond this value, indicating the cutoff by a vertical line in Fig.~\ref{fig:1}. The shape of the spectrum for $k\gtrsim \beta$ is a subject of ongoing research and is beyond the scope of this work. 

In Fig.~\ref{fig:1}, we show the power-law–integrated (PLI) sensitivity curves for various upcoming gravitational-wave (GW) missions, following Ref.~\cite{Thrane:2013oya}. The shaded regions indicate the projected sensitivity reaches of the respective experiments. These curves assume that the GW spectrum from a first-order phase transition can be approximated by a power law, $\Omega_{\rm GW}\propto f^{b}$, where $b$ is the spectral index. A theoretical prediction lying within the shaded region corresponds to a signal detectable with high signal-to-noise ratio (SNR). While alternative approaches, such as peaked-integrated sensitivity curves (PISCs), may better characterize certain phase transition–induced GW spectra~\cite{Schmitz:2020syl}, we show the PLI curves for illustration and perform a dedicated SNR analysis next. 

\medskip

In the same diagram we compare the GW signals as in Eq.~(\ref{27},\ref{30}) with respect to the PLI curves. It has been shown in the literature that for super-horizon modes the GW power spectrum sourced by a FOPT exhibits a universal scaling, $\Omega_{\rm GW}h^2\propto f^{3}$ (see \cite{Cai_2020}). In the deep-IR regime, i.e. at frequencies corresponding to timescales longer than a Hubble time—the spectrum therefore follows the $f^{3}$ scaling due to causality~\cite{PhysRevD.79.083519} rather than the $f^{1}$ behavior discussed for sub-horizon particle-production sources. To account for this, we impose an IR cutoff at the Hubble frequency, $f_{IR} = H(T_*)$, and adopt the $f^{3}$ scaling for $f<f_{\rm IR}$. We stress that this cutoff is not a sharp, model-independent boundary: the precise transition point in the IR must be checked numerically and can vary with the details of the phase transition.\footnote{See Eq.~(33) of Ref.~\cite{Inomata:2024rkt} for analytical expression.} Note we impose a similar IR cutoff for GWs coming from bubble collision as well.

\medskip

Also for GW contribution coming from particle production, we have placed the UV cutoff at $k/\beta = 0.75$ as discussed earlier in this section. We have found for a given $\alpha,\ T_*$ - bigger values of $\beta/H$ corresponds to a higher $\Omega_{GW}^{pp}h^2$ over $\Omega_{GW}^{coll}h^2$. Also note, for different phase transition temperatures, the peak frequency shifts and hence different detectors can be used to probe the GW spectrum. For example in Fig-[\ref{fig:1}], for BP1 with $T_* = 10^{-5}$ TeV, one can find the signal detectable in various Pulsar Timing Arrays (PTA) such as EPTA or NanoGrav whereas for BP2 with $T_* = 1$ TeV one detect the GW signals using LISA, BBO or $\mu$ARES experiments. Also Fig.~\ref{fig:1} hints that for certain parameters and detectors, the GW signal coming from particle production can be distinguished form its bubble collision counterpart. Although the particle production contribution is mostly subsumed within the bubble collision spectrum, its distinct spectral shape and benchmark dependence can reveal regions of parameter space where it can be uniquely identified. This will be crucial for shedding light on some reheating mechanisms in dark or visible sector or both. To further investigate the scenario in detail, we examine the Signal-to-Noise ratios (SNRs) for different detectors for both of the GW production mechanisms over the parameter space in the next section.

\subsection{Signal-to-Noise-Ratio (SNR)}

\noindent Interferometers are instruments measuring displacements in terms of dimensionless strain-noise denoted as $h_{\text{GW}}(f)$. This strain-noise is related to the amplitude of GWs passing by, usually converted into an energy density quantity, expressed by the formula,

\begin{equation}
    \Omega_{\text{exp}}(f)h^2 = \frac{2\pi^2 f^2}{3\mathcal{H}_0^2} h_{\text{GW}}(f)^2h^2.
\end{equation}

\medskip

\noindent Here, $\mathcal{H}_0$ represents the present-day Hubble rate ($\mathcal{H}_0 = h \times 100 \frac{\text{km/s}}{\text{Mpc}}$). To investigate the likelihood of detecting the primordial GW background, we calculate the signal-to-noise ratio (SNR) using the experimental sensitivity for the noise curves $\Omega_{\text{exp}}(f)h^2$, either given or projected for future experiments. The SNR is got using the following expressions,

\begin{equation}\label{eq:SNR}
    \text{SNR} \equiv \sqrt{\tau \int_{f_{\text{min}}}^{f_{\text{max}}} df \left(\frac{\Omega_{\text{GW}}(f)h^2}{\Omega_{\text{exp}}(f)h^2}\right)^2}.
\end{equation}

\medskip

\noindent where we use $h = 0.67$ and an observation time of $\tau = 4$ years for all GW detectors except the PTA and SKA based GW detectors for which we take $\tau = 20$ years. A detection threshold of SNR $\geq 10$ is applied. It is important to note that this formula for SNR calculation is got under the weak signal approximation, where the GW signal is much smaller than the instrumental noise~\cite{Allen:1997ad}. While it can at certain times overestimate the true SNR for strong signals, we adopt it here for both weak and strong GW signals for simplicity. 

\medskip

\subsubsection{Gravitational Wave Detectors} 
In the GW spectrum plots, we present the sensitivity curves for different ongoing and future GW experiments. They can be grouped as: 
\begin{itemize}
    \item \textbf{Ground-based interferometers:} These detectors, such as \textsc{LIGO}/\textsc{VIRGO}/\textsc{KAGRA} (LVK)\cite{LIGOScientific:2016aoc,LIGOScientific:2016sjg,LIGOScientific:2017bnn,LIGOScientific:2017vox,LIGOScientific:2017ycc,LIGOScientific:2017vwq,KAGRA:2013rdx}, a\textsc{LIGO}/a\textsc{VIRGO} \cite{LIGOScientific:2014pky,VIRGO:2014yos,LIGOScientific:2019lzm}, \textsc{AION} \cite{Badurina:2021rgt,Graham:2016plp,Graham:2017pmn,Badurina:2019hst}, \textsc{Einstein Telescope (ET)} \cite{Punturo:2010zz,Hild:2010id}, and \textsc{Cosmic Explorer (CE)} \cite{LIGOScientific:2016wof,Reitze:2019iox}, use interferometric techniques on the Earth's surface to detect gravitational waves.
    
    \item \textbf{Space-based interferometers:} Space-based detectors like \textsc{LISA} \cite{Baker:2019nia}, \textsc{BBO} \cite{Crowder:2005nr,Corbin:2005ny,Cutler:2009qv}, \textsc{DECIGO}, \textsc{U-DECIGO} \cite{Seto:2001qf,Yagi:2011wg}, \textsc{AEDGE} \cite{AEDGE:2019nxb,Badurina:2021rgt}, and \textsc{$\mu$-ARES} \cite{Sesana:2019vho} are designed to detect gravitational waves from space, offering different advantages over ground-based counterparts.
    
    \item \textbf{Recasts of star surveys:} Surveys such as \textsc{GAIA}/\textsc{THEIA} \cite{Garcia-Bellido:2021zgu} utilize astrometric data from stars to indirectly infer the presence of gravitational waves.
    
    \item \textbf{Pulsar timing arrays (PTA):} PTA experiments like \textsc{SKA} \cite{Carilli:2004nx,Janssen:2014dka,Weltman:2018zrl}, \textsc{EPTA} \cite{EPTA:2015qep,EPTA:2015gke}, and \textsc{NANOGRAV} \cite{NANOGRAV:2018hou,Aggarwal:2018mgp,NANOGrav:2020bcs} use precise timing measurements of such pulsars to measure gravitational wave signatures.
\end{itemize}

\subsection{Numerical Results}
\begin{figure}[H]
    \centering
    \includegraphics[width=0.7\linewidth]{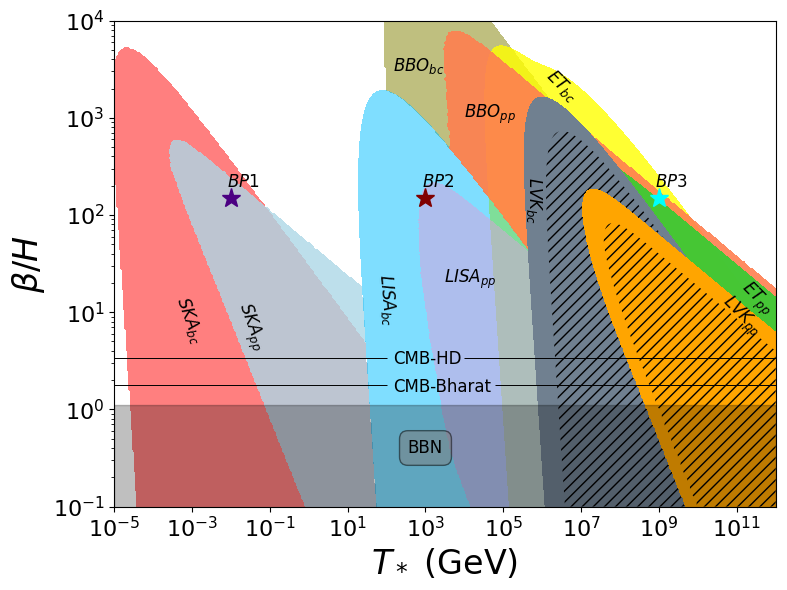}
    \caption{\it  The parameter reach of the present and future GW experiment network. In each shaded region, corresponding experiment (see labels) detects the GW signal coming from a FOPT with SNR $\geq 10$. On top of this plane, the three benchmark points BP1, BP2 and BP3, taken in Fig~\ref{fig:1} is shown.   We have used $\alpha = 10$ and $\kappa = 1$. The subscript `bc' stands for `bubble collision' and `pp' for
    `particle production'. The BBN bound (horizontal gray band) rules out the region $\beta/H < 1.1$. The hatched gray and orange regions have been ruled out from LVK $\mathcal{O}(3)$ data. }
    \label{fig:2}
\end{figure}

\noindent We plot the Signal-to-Noise ratio for different detectors in the $T_*$ vs $\beta/H$ plane considering both bubble collision and particle production mechanisms as sources in Fig-\ref{fig:2}. We have checked numerically for large $\alpha$ the SNR values remain more or less constant and hence choose one benchmark for $\alpha$ only.  

We remark the following observations:
\begin{enumerate}
    \item \textbf{Single detector distinguishability :} There are regions of parameter space, for example in SKA with $\beta/H = 25,\ T_*=10$ GeV where one can detect GW from particle production but not from bubble collision and vice-versa.
    \item \textbf{Inter-detector distinguishability :}
    There are regions of parameter space for a set of detectors, for example, LISA and SKA with $\beta/H = 10,\ T_* = 100$ GeV, where signal coming from particle production can be detected in SKA and signal coming from bubble collision can be detected in LISA.
\end{enumerate}

\noindent It is worth emphasizing that an SGWB source yielding a given signal-to-noise ratio (SNR) in a detector merely indicates that its signal would be present in the data with a certain statistical significance. Nevertheless, the SNR criterion alone does not guarantee an actual detection. In particular, astrophysical foregrounds arising from unresolved individual sources may obscure even SGWB signals with comparatively high SNR values. Furthermore, for signal-dominated experiments where the noise curve must be determined together with the signals, the SNR evaluation does not include the uncertainties on the noise reconstruction. A precise assessment of the detection capabilities would therefore require running the full SGWB reconstruction pipeline specific to each experiment and Fisher matrix analysis which is beyond the scope of the present analysis given the complexity. Here we restrict ourselves to the SNR estimates and leave a more quantitative evaluation of detection prospects for future work.

\subsection{Bounds from BBN and CMB on dark radiation}
\noindent The gravitational wave energy density has to obey certain observational bounds, for instance, should be smaller than the bound on dark radiation parameterized as the number of relativistic neutrino species $\Delta N_{\rm eff}$~\cite{Luo:2020fdt,Maggiore:1999vm},
\begin{equation}
    \int_{f_{\rm min}}^{\infty}\frac{df}{f}\Omega_{\rm GW}(f)h^2\leq5.6\times10^{-6}\Delta N_{\rm eff}
\end{equation}
We typically do not consider the frequency dependence and neglect it, for this bound and set $\Omega_{\rm GW}\leq 5.6\times10^{-6}\Delta N_{\rm eff}$ for the GW spectra that we calculate. BBN puts a bound on $\Delta N_{\rm eff}^{\rm BBN}\simeq 0.4$~\cite{Cyburt:2015mya}. Planck plus BAO observations set the bound $\Delta N_{\rm eff}^{\rm Plack+BAO}\simeq0.28$~\cite{Planck:2018vyg}. Future projected bounds from future experiments are $\Delta N_{\rm eff}^{\rm Proj.}=0.014$ for CMB-HD~\cite{CMB-HD:2022bsz}, $\Delta N_{\rm eff}^{\rm Proj.}=0.05$ for CMB-Bharat~\cite{CMB_Bharat}, $\Delta N_{\rm eff}^{\rm Proj.}=0.06$ for CMB Stage IV~\cite{Abazajian:2019oqj} and NASA's PICO mission~\cite{Alvarez:2019rhd}, $\Delta N_{\rm eff}^{\rm Proj.}\lesssim0.12$ for CORE~\cite{CORE:2017oje}, the South Pole Telescope~\cite{SPT-3G:2014dbx} and Simons observatory detectors~\cite{SimonsObservatory:2018koc}.

\medskip

\section{Cosmological relics from Particle Production from Bubbles}
\label{IV}
 \noindent We consider a first-order phase transition in a dark (hidden) sector in which a scalar field $\phi$ tunnels from a metastable false vacuum with $\langle\phi\rangle = 0$ to a true vacuum with $\langle\phi\rangle = v_\phi$. For illustration, we parameterize the released latent heat as $\Delta V = \lambda v_\phi^4$.~\footnote{The parameter $\lambda$ is different from the scalar quartic coupling $\lambda_\phi$ and in general contains temperature contributions. For supercool phase transitions with $\alpha\gg 1$, the thermal contributions to the latent heat can be neglected.} The parameter $\lambda$ will reduce to the scalar quartic coupling for a $\phi^4$ theory but in general, the potential may have quadratic and cubic terms and $\lambda$ may be different from the quartic coupling. In what follows, we study the couplings of different particle species to $\phi$ and their production via the mechanism described in Sec.~\ref{II}, assessing their viability in generating the required relic abundances and the associated gravitational-wave signals discussed in Sec.~\ref{III}.

\subsection{Non-thermal Production of Right-handed Neutrino }

\noindent Right-handed neutrinos (RHNs) play a crucial role in explaining the Standard Model neutrino masses and the observed baryon asymmetry via leptogenesis. Additionally, they can serve as natural dark-matter candidates. In this section, we consider RHNs coupled to the bubble walls so that bubble collisions generate an initial frozen-in RHN population, and we explore the parameter space relevant for each scenario. Note, in all subsequent cases, the masses of the RHNs lie well beyond the typical phase transition temperatures and hence at production, the thermal population always remains negligible.\\

We focus on a heavy RHN species with mass $M_N$ coupled to a scalar field $\phi$ undergoing the phase transition. The interaction Lagrangian is
\begin{equation}
    -\mathcal{L}_I \supset M_N\bar{N}^cN + y_\phi\phi\bar{N}^cN +\ h.c.,
\end{equation}
where $y_\phi$ denotes the Yukawa coupling to $\phi$. If the bubbles of background field $\phi$ achieve runaway behavior, releasing vacuum energy $\Delta V = \lambda v_\phi^4$ as discussed earlier, then the produced RHN number density from bubble collision, given by Eq.~(\ref{10}) is (see~\cite{cataldi2024leptogenesisbubblecollisions}) 
\begin{equation}
    n_{N} \simeq 1.6 y_\phi^2\frac{\beta}{H}\left(\frac{30(1+\alpha)\lambda}{\pi^2\alpha} \right) ^{1/2}\frac{v_\phi^4}{M_{pl}}\ln{\left(\frac{2E_{max}}{M_{N}}\right)},
    \label{44}
\end{equation}
with 
\begin{equation}
     E_{max} = \frac{\gamma_{max}}{l_{w0}} \sim 0.7 \left(\frac{\alpha}{(1+\alpha)\lambda}\right)^{1/2} \frac{M_{pl}}{\beta/H}.
     \label{45}
\end{equation}
\medskip

\noindent Here the Planck mass $M_{pl} = 1.22\times 10^{19}$ GeV and $\gamma_{max}$ is the maximum value of the Lorentz factor achieved by the bubble wall. Note due to the scalar self coupling, $\phi$ particles would also be produced from the phase transition. These can later decay into the $N$ states and change the yield significantly. The number density of $\phi$ particles produced from the process $\phi^*\rightarrow \phi\phi$ is~\cite{cataldi2024leptogenesisbubblecollisions} 
\begin{equation}
    n_\phi \approx 0.64 \lambda_\phi^2\frac{\beta}{H}\left(\frac{30(1+\alpha)\lambda}{\pi^2\alpha} \right) ^{1/2}\frac{v_\phi^4}{M_{pl}}
\end{equation}
and from the process $\phi^*\rightarrow 3\phi$ is 
\begin{equation}
    n_\phi \approx \frac{\lambda_\phi^2}{48\pi^2} \frac{\beta}{H}\left(\frac{30(1+\alpha)\lambda}{\pi^2\alpha} \right) ^{1/2}\frac{v_\phi^4}{M_{pl}}\ln\left(\frac{2 E_{max}}{3 m_\phi}\right).
    \label{fortyfour}
\end{equation}
\medskip
Here $\lambda_\phi$ denotes the scalar quartic coupling, and we assume $m_\phi \simeq v_\phi$. For nonrelativistic $\phi$ production with $E_\phi \simeq m_\phi$, the two-body decay channel dominates, while for relativistic production with $E_\phi \gg m_\phi$, the three-body channel becomes dominant. In both regimes, Eqs.~(\ref{44}-\ref{fortyfour}) show that the ratio of produced RHNs to produced $\phi$ quanta from bubble collisions is approximately
\begin{equation}
    \frac{n_N}{n_\phi} \approx \left(\frac{y_\phi}{\lambda_\phi}\right)^2 .
    \label{RHN-Phi}
\end{equation}
\medskip
Thus, taking $y_\phi \gtrsim 10\lambda_\phi$ ensures that the RHN yield is dominated by direct production from bubble collisions rather than from subsequent $\phi$ decays.

\vspace{0.2cm}

With this assumption, the resulting RHN yield is given by: 
\begin{equation}
    \mathcal{Y}_N = \frac{n_N}{s} \sim 4y_\phi^2\frac{\beta}{H}\left(\frac{\pi^2\alpha}{30(1+\alpha)g_*\lambda}\right)^{1/4}\frac{v_\phi}{M_{pl}}\ln{\left(\frac{2E_{max}}{M_{N}}\right)}.
    \label{46}
\end{equation}
Here $g_*$ denotes the effective number of relativistic degrees of freedom in the surrounding plasma. The RHN mass in the broken phase is
\begin{equation*}
    M_N = M_N^0 + y_\phi v_\phi,
\end{equation*}
where $M_N^0$ is the bare mass and $y_N v_\phi$ is the contribution from the vacuum expectation value of $\phi$. To express the yield fully in terms of the phase transition parameters, we use : 
\begin{equation*}
    v_\phi = \left(\frac{\pi^2g_*\alpha}{30\lambda}\right)^{1/4}T_*.
\end{equation*} 
Substituting this relation in Eqn \eqref{46}, the RHN yield becomes 
\begin{equation}
    \mathcal{Y}_N \approx 4y_\phi^2\frac{\beta}{H}\left(\frac{\pi^2\alpha}{30\lambda}\right)^{1/2}(1+\alpha)^{-1/4}\frac{T_*}{M_{pl}}\ln{\left(\frac{2E_{max}}{M_{N}}\right)}.
    \label{fortyfive}
\end{equation}

\noindent Assuming this is the Freeze in yield, the RHN relic abundance today is given by:
\begin{equation}
    \Omega_N = \left.\frac{\rho_N}{\rho_c}\right|_0 = \frac{M_N \mathcal{Y}_Ns_0}{3M_{pl}^2H_0^2} \implies \Omega_N h^2 \approx 1.89\times10^3\ \mathcal{Y}_N\frac{M_N}{GeV},
    \label{48}
\end{equation}

\vspace{0.2cm}

\noindent where we have taken the entropy density today to be $s_0 = 7.04n_\gamma$ and $n_\gamma = 400\ cm^{-3}$. With these expressions in hand, we evaluate the RHN relic abundance for different values of $M_N$. Throughout, we take $\lambda \sim \mathcal{O}(1)$. Since the yield scales as $\mathcal{Y}_N \propto \lambda^{-1/2}$, its variation is modest. 

\vspace{0.3cm}

As mentioned earlier, for a $\phi^4$ potential, $\lambda$ would be the quartic self coupling $\lambda_\phi$. Then choosing $\lambda_\phi \sim 0.1$—consistent with the requirement $y_\phi \gtrsim \mathcal{O}(10)\lambda_\phi$—changes the yield only by a factor of $10^{1/2} \simeq 3.1$. For more complex potentials, value of $\lambda$ needs to be determined for each case, and can be different from $ \mathcal{O}(1)$. We do not deal with details of model-building in the present paper but will be taken in future. 

\subsection{DM formation and its GW probe}

\noindent 
As noted in the introduction, observations indicate the presence of a non-luminous matter component comprising roughly $26\%$ of the Universe’s energy density — also known as dark matter (DM)~\cite{Zwicky:1933gu}. Since the Standard Model lacks a viable DM candidate, various beyond-the-Standard-Model (BSM) scenarios have been proposed~\cite{Cirelli:2024ssz} in order to explain the DM. In this section, we explore whether the RHNs introduced above—when rendered stable—can account for the observed DM relic abundance.

\medskip

 This can be achieved by imposing an additional stabilizing symmetry, for example a simple $\mathcal{Z}_2$ under which the lightest RHN is odd while all Standard Model fields are even\footnote{Note that imposing an additional $\mathcal{Z}_2$ symmetry forbids the participation of RHNs in the seesaw mechanism and leptogenesis, as discussed in the following sections. Alternatively, allowing the RHNs to have sufficiently small couplings renders them stable on cosmological timescales, enabling them to serve as dark matter candidates while still realizing the seesaw mechanism.}. In this setup, the lightest RHN cannot decay into SM states and is therefore stable: 
 \begin{equation}
     SM\xrightarrow{\mathcal{Z}_2} SM\ ; \quad N_1\xrightarrow{\mathcal{Z}_2}-N_1.
 \end{equation}
 Protected by the $\mathcal{Z}_2$ symmetry, we now explore the parameter space relevant for such stable RHN dark-matter scenario. 

\medskip

We first demonstrate that the RHN relic abundance $\Omega_N$, which in this scenario constitutes the total dark-matter density, $\Omega_N = \Omega_{\rm DM}$, can reproduce the observed value\footnote{Note that in this case the RHNs are produced via bubble wall collisions, which is qualitatively different from the conventional freeze-in dark matter scenario, where freeze-in happens from thermal bath.} $\Omega_{\rm DM} h^2 \simeq 0.12$ for appropriate choices of the RHN mass and coupling, $(M_N, y_{\phi})$, as obtained from Eq.~(\ref{48}). For the numerical analysis, we adopt the phase transition benchmark BP2 of Fig.~\ref{fig:1}, for which the associated GW signal lies within the prospective sensitivity of upcoming detectors such as LISA and BBO.

\medskip

\begin{figure}[H]
    \centering
    \includegraphics[width=0.48\linewidth]{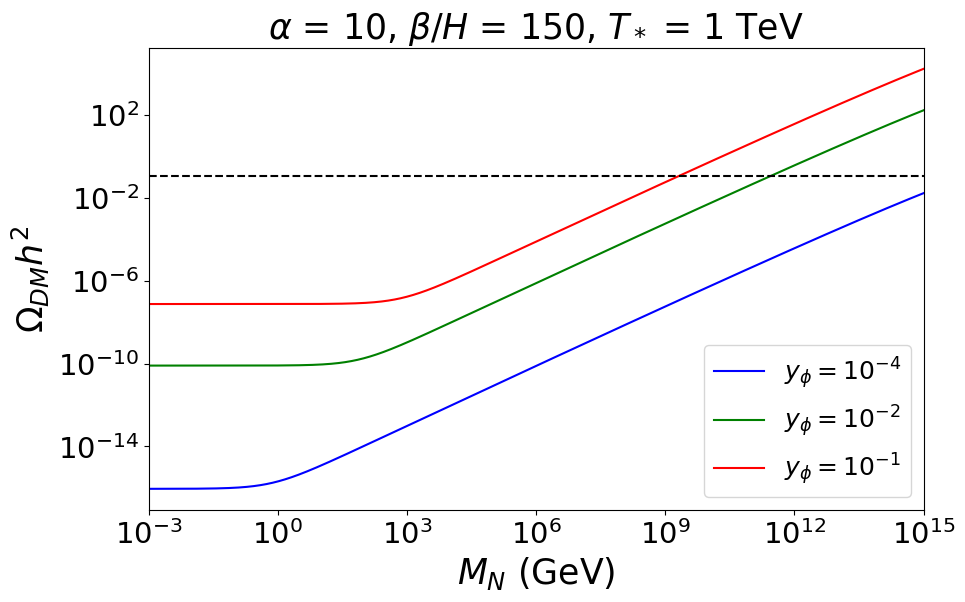}
    \includegraphics[width = 0.48\linewidth]{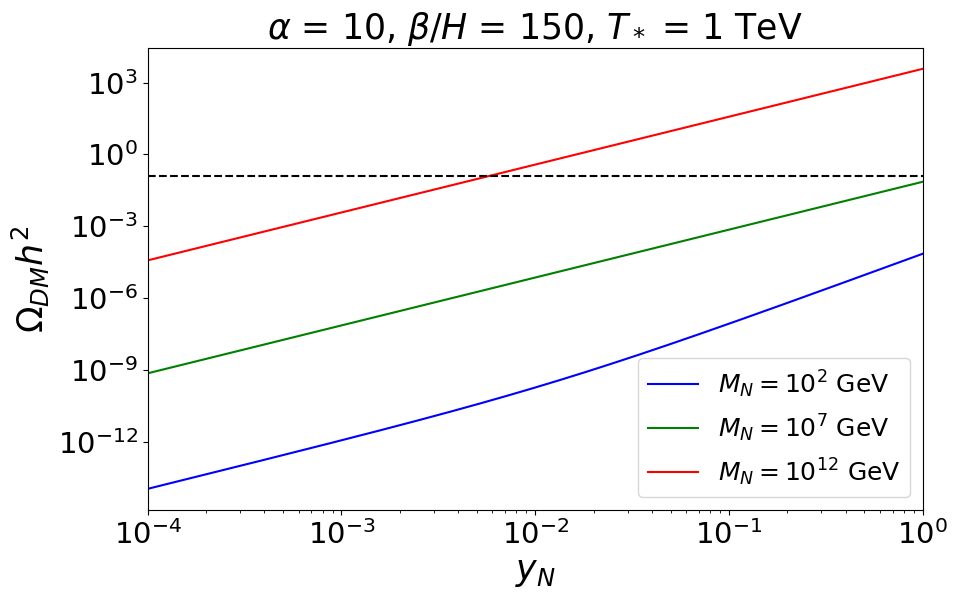}
    \caption{\it  For PT parameter \textbf{BP-2} (a) DM Relic vs RHN mass : plotted for three different benchmark points of the coupling strength (b) DM relic vs RHN coupling : again plotted for three different benchmark points of the RHN mass. In both cases $\lambda = \mathcal{O}(10^{-1})$ is taken.}
    \label{fig:3}
\end{figure}

\noindent As discussed in the previous section, it is important to ensure that the $\phi$ particles produced via the quartic self-coupling $\lambda_\phi$ during the phase transition do not contribute significantly to the dark-matter abundance. Since $\phi$ is unstable , it cannot constitute DM itself. Depending on the specific form of the potential, the parameter $\lambda$ must be adjusted\footnote{The parameter $\lambda$ in general is different from $\lambda_\phi$ and needs to be calculated from the potential.} such that $y_{\phi} \sim 10\lambda_\phi$ (see Eq.~\ref{RHN-Phi}), ensuring that the contribution of $\phi$ to the RHN yield remains negligible. 

\medskip

\begin{figure}[H]
    \centering
    \includegraphics[width=0.7\linewidth]{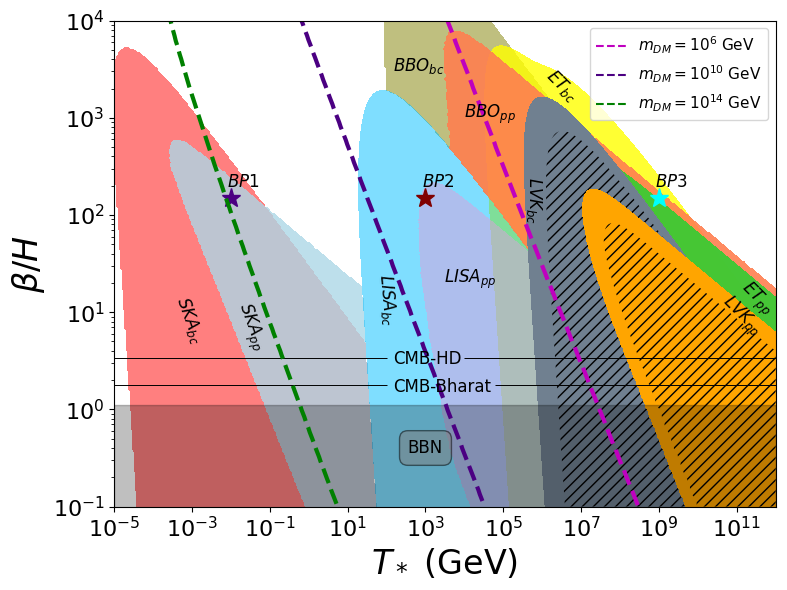}
    \caption{\it  Contours of $\Omega_{DM}h^2 = 0.12$ plotted for three different benchmark masses $M_N \equiv m_{DM} = 10^6,10^{10}$  and $10^{12}$ GeV with $y_{\phi} = 0.5$. In the background, regions with SNR $\geq 10$ shown for different detectors for GW signal coming from both bubble collision and particle production. $\lambda\sim \mathcal{O}(1)$ taken. The hatched gray and orange regions have been ruled out from LVK $\mathcal{O}(3)$ data.}
    \label{fig:5}
\end{figure}

With this in mind, we observe in Fig.~\ref{fig:3} that in both panels the red curves intersect the $\Omega_{\rm DM} h^{2} = 0.12$ line. This shows that for sufficiently large masses, $M_N \geq 10^{9}$ GeV, and couplings, $y_{\phi}\geq 0.01$, the RHN relic abundance matches the observed DM density. A complete parameter scan is taken up in Fig.~\ref{fig:5}. For small values of $M_N$, we have $M_N\lesssim y_{\phi}v_\phi$, and hence the mass essentially becomes constant ($\sim y_{\phi}v_\phi$) for $M_N\lesssim 1$ GeV. This leads to the nearly constant behavior of $\Omega_{\rm DM}h^2$ for $M_N\lesssim1$ GeV, as seen in the left panel of Fig.~\ref{fig:3}. However, several assumptions entering Eq.~(\ref{46}) break down in the regime $M_N\lesssim v_\phi$, and results in this region should not be considered reliable\footnote{If the particle gains mass mainly from the FOPT, i.e. $m\sim v_\phi$, its Compton wavelength $(\sim m^{-1})$ can exceed the thickness of the highly boosted bubble walls $(l_w)$ at collisions. In this regime, the particle probes both vacua on either side of the bubble wall simultaneously, and the notion of a well-defined particle state during the collision breaks down, see Ref.~\cite{Mansour:2023fwj} for a detailed discussion. }. Motivated by these considerations, we now perform a parameter scan in the $M_N-y_{\phi}$ plane.

\begin{figure}[H]
    \centering
    \includegraphics[width=0.49\linewidth]{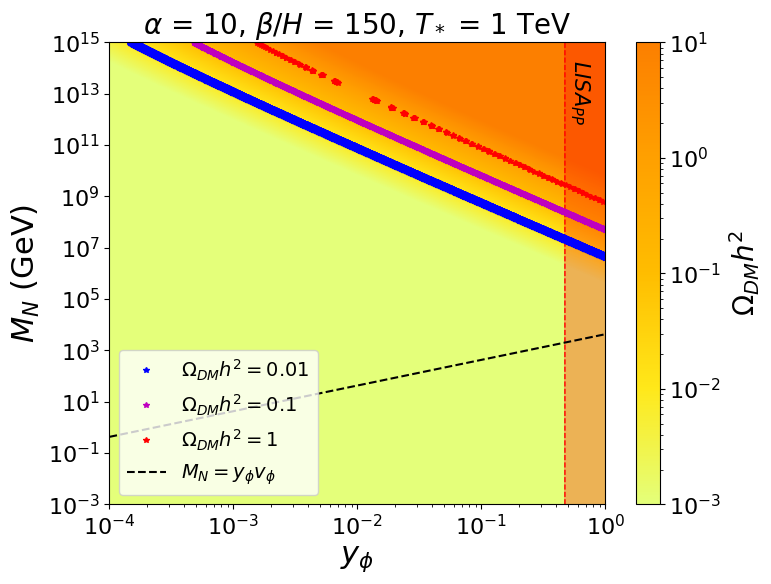}
    \includegraphics[width=0.49\linewidth]{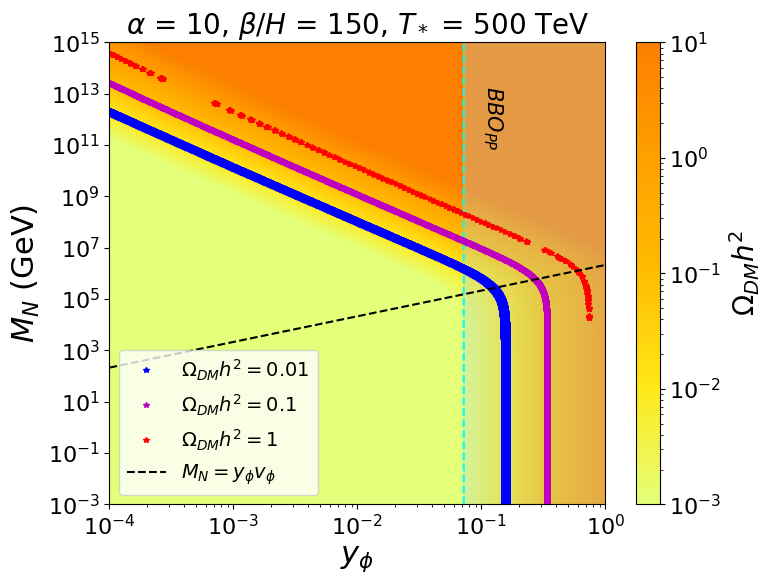}
    \caption{\it DM relic density in the $(m_{\rm DM}, y_{\rm DM})$ plane for two phase transition benchmarks with $\mathcal{O}(1)$ values of $\lambda$. \textbf{Left:} BP2, where bubble-collision GWs are detectable in LISA and BBO with $\mathrm{SNR}>10$ for the entire plane. \textbf{Right:} $\alpha=10$, $\beta/H=150$, and $T_*=500~\mathrm{TeV}$, with bubble-collision GWs detectable in BBO for the entire plane. Upper triangular regions are excluded due to over-closure of the universe. Vertical dotted lines indicate $\mathrm{SNR}=10$ for particle-production GWs for different detectors (see label); regions to the right of these lines yield $\mathrm{SNR}>10$. Values of $\kappa$ for the RHN production GWs is given by Eq.~\ref{fiftyone}. }
    \label{fig:4}
\end{figure}
\noindent Figure~\ref{fig:4} shows that the observed DM relic abundance can be obtained for $y_{\phi}>10^{-3}$ with $M_N\in [10^4,10^{12}]$ TeV for $T_* = 1$ TeV and for $y_{\phi}\in [10^{-4},0.2]$ with $M_N\in [10^{3},10^{10}]$ TeV for $T_* = 500$ TeV. Part of this parameter space is also within the reach of future GW detectors such as LISA, BBO. We additionally show the line $M_N = y_{\phi}v_\phi$, below which the RHNs receive most of their mass from the phase transition (via the vev contribution). As noted earlier, results below this line are not reliable within the particle-production formalism adopted in this work. Finally, we emphasize that the quoted SNR values for the GW signals sourced by particle production depend on the efficiency factor $\kappa$ (see Eq.~\ref{36}) which in turn depends on the details of the particle production mechanism via the relation Eq.~(\ref{38}). For our scenario of RHN as DM production via Yukawa coupling, this is calculated to be
\medskip
\begin{equation}
    \kappa \sim \frac{6 y_{\phi}^2}{\lambda\pi^3}\log[\dots]\left(1+\mathcal{O}(M_{pl}^{-1})\right),
    \label{fiftyone}
\end{equation}
where the logarithmic factor is identical to that appearing in Eq.~(\ref{12}) and remains approximately constant—between 6 and 60—for the values of $\beta/H,\ \alpha$ etc. in the range considered in sec.~\ref{II}. For the same set of parameters, the SNRs of bubble–collision–induced GWs remain constant across each plot. In particular, for $T_* = 1$ TeV, bubble collision SNRs for BBO and LISA are 1321 and 18 respectively, whereas for $T_* = 500$ TeV, bubble collision SNRs for BBO and LISA respectively are $10^6$ and 0.03 respectively. Therefore, Fig.~\ref{fig:4} illustrates that if the produced RHNs are stable, they can constitute viable dark matter candidates, with the additional prospect of indirect detection via their associated GW signatures. 

\medskip

\noindent To further investigate this observation, we perform a detailed parameter-space scan in the $\beta/H-T_*$ plane, highlighting the detection regions relevant for current and future GW experiments.
We also overlay contours of $\Omega_{DM}h^2 = 0.12$ corresponding to three benchmark RHN mass values to highlight the parameter space consistent with both dark matter abundance and GW observability.

We see in Fig.~\ref{fig:5}, The GW signals coming from the FOPT event that produces correct relic abundance for the stable RHN dark matter can be successfully detected in multiple current and upcoming detectors. For mass as low as $M_N = 10^6$ GeV, GW signal can be detected in the BBO and future LVK detectors for both particle production and bubble collision mechanisms whereas ET can detect only bubble collision GWs in this region. Part of the parameter space is already ruled out by LVK-$\mathcal{O}(3)$ data and further parameter space can be probed with future LVK observations. On the other hand, for $M_N = 10^{10}$ GeV, the corresponding GW signals coming from the bubble collisions can be detected in LISA and GW signals coming from particle production can be detected in SKA. Similarly, for very high mass $M_N = 10^{14}$ GeV, the GW signals for both particle production and bubble collision can be detected in SKA. Note as was discussed in sec~\ref{III} and can be seen in Eq.~(\ref{27}\&\ref{30}), the corresponding GW signals coming from particle production and bubble collision has different scaling properties with frequency. This is particularly prominent in the low frequency regions of a given GW spectrum, and hence probing both contributions in separate detectors for the same PT parameters re-enforces the validity of the prediction.

\subsection{Non-thermal Leptogenesis from Bubble collisions}
\noindent As noted in the introduction, the observed baryon asymmetry and the origin of neutrino masses remain among the most pressing open questions in particle physics, and both can be addressed via leptogenesis. In this section, we explore the possibility of realizing leptogenesis through the production of non-thermal RHNs during a FOPT in this  model \footnote{Leptogenesis in the context of strong first order phase transition has been studied in the literature but in a different context \cite{Dasgupta:2022isg,Ghosh:2025non,Athron:2025pog,Ai:2025vfi,Cataldi:2025nac}.}. We also investigate the corresponding gravitational-wave signals arising from both bubble-collision and particle-production mechanisms, analogous to the analysis performed for dark matter in the previous section.

\subsubsection{Type-I Seesaw}
The well-known seesaw mechanism provides a natural explanation for the tiny active neutrino masses via extending the SM with heavy fermion singlets (right-handed neutrinos N) at high energy scales. The light neutrino masses emerge as inversely proportional to the masses of these heavy states, so larger the heavy right-handed neutrino masses, the smaller the SM neutrino mass, (therefore the ``seesaw")
\begin{equation}
    m_{\nu}\propto\frac{v ^2}{M_N}
    \label{fiftwo}
\end{equation}
where $M_N$ is the mass of the heavy seesaw state and $v$ is the electroweak scale vacuum expectation value (vev). For this work we start with a conventional Type I Seesaw\cite{Minkowski:1977sc,gellmann2013complexspinorsunifiedtheories,Glashow:1979nm},with three right handed neutrinos $N$ 
\begin{equation}
    \mathcal{L} \supset \lambda_N \bar{L}(i\sigma_2)H^\dagger N + \frac{1}{2}M_N\bar{N}^cN + h.c.,
    \label{fiftytwo}
\end{equation}
where $L$ is the SM lepton doublet, $H$ is the $SU(2)_L$ Higgs doublet and $\sigma_2$ is the second Pauli matrix. Without any loss of generality we can take the RHN mass matrix to be diagonal in this basis 
\begin{equation}
    M_N = \text{diag}(M_1,M_2,M_3).
\end{equation}
After Electroweak Symmetry Breaking (EWSB), Higgs acquires a vev $v = 174$ GeV. Integrating out the heavy RHNs, one generates SM neutrino masses which to the leading order in Seesaw expansion reads 
\begin{equation}
    m_\nu  = -m_D\cdot M_N^{-1} m_D^t\ ; \quad \text{with,}\ \ m_D \equiv \lambda v \ll M_N.
\end{equation}

To go to a basis where the charged lepton mass matrix is diagonal, we adopt the Casas-Ibarra parametrization~\cite{Casas_2001} and write the lepton interaction matrix as 
\begin{equation}
    \lambda_N = \frac{1}{v}\cdot M_N^{1/2}\cdot R\cdot m_\nu^{1/2}\cdot U_{PMNS}^\dagger,
    \label{54}
\end{equation}
where $U_{PMNS}$ is the leptonic equivalent of the CKM matrix. $R$ describes the mixing and
CP-violation in the RHN sector and is expressed as a complex, orthogonal matrix that reads
\begin{equation}
    R \equiv \text{diag}(\pm 1,\pm 1,\pm 1)\cdot R_{23}(\theta_{23})\cdot R_{13}(\theta_{13})\cdot R_{12}(\theta_{12}),
\end{equation}
where $R_{ij}$ are $2\times2$ rotation matrices with angle of rotation $\theta_{ij}$. The lightest RHN $N_1$ has the total decay width to the SM leptons given by
\begin{equation}
    \Gamma_1\equiv \Gamma(N_1\rightarrow LH,\bar{L}H^\dagger) = \frac{|\lambda_N^\dagger\lambda_N|_{11}}{8\pi}M_1.
    \label{56}
\end{equation}
For relativistic RHN this rate gets suppressed by the Lorentz factor $\gamma \sim E_1/M_1$, where $E_1$ is the energy of the relativistic RHN. The effective neutrino mass mediated by $N_1$ is defined as - 
\begin{equation}
    \tilde{m}_1 \equiv \frac{|\lambda_N^\dagger\lambda_N|_{11}v^2}{M_1} = \sum_i m_i|R_{1i}|^2
    \label{57}
\end{equation}
where the last equality follows from Eq.~(\ref{54}). This effective mass appears in when comparing the decay rate with the characteristic time scale of the cosmological expansion.

\begin{figure}[H]
    \centering
    \includegraphics[width=0.8\linewidth]{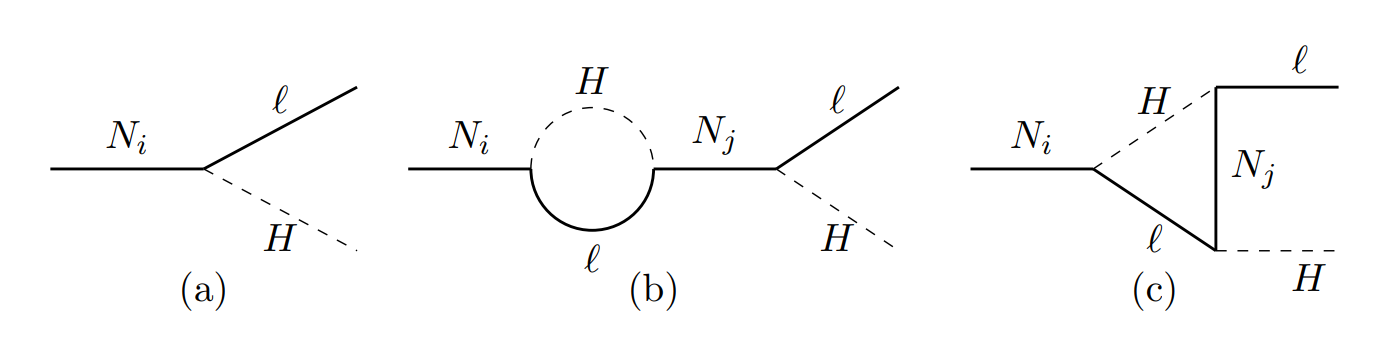}
    \caption{\it RHN decay into SM $N\rightarrow LH$ contributing to CP violation: (a) tree level decay (b) self energy contribution (c) vertex diagrams }
    \label{fig:six}
\end{figure}

The lepton asymmetry will be generated in RHN decays $N\rightarrow L H$ via interference between
tree-level and one-loop diagrams. The
relevant decay processes are depicted in Fig~\ref{fig:six}.
The amount of CP violation is quantified by the CP violation parameter $\epsilon_1$ defined as
\begin{equation}
    \epsilon_1 = \frac{\Gamma(N_1\rightarrow \bar{L}H) - \Gamma(N_1\rightarrow LH^\dagger)}{\Gamma(N_1\rightarrow \bar{L}H) + \Gamma(N_1\rightarrow LH^\dagger)},
\end{equation}
which using Eq.~(\ref{54},\ref{56}) simplifies to 
\begin{equation}
    \epsilon_1 = \sum_{i\neq 1} \frac{3}{16\pi}\frac{M_1}{M_i}\frac{\text{Im}\left((\lambda\lambda^\dagger)^2_{1i}\right)}{|\lambda\lambda^\dagger|_{11}} = \frac{3}{16\pi}\frac{M_1}{v^2}\frac{\sum_i m_i^2\text{Im}(R_{1i}^2)}{\sum_j m_j |R_{ij}|^2} \leq \epsilon_{max}.
    \label{CP1}
\end{equation}
Here $\epsilon_{max}$ is related to the Davidson-Ibarra bound~\cite{Davidson_2002} and is given by 
\begin{equation}
    \epsilon_{max} = \frac{3}{16\pi}\frac{M_1}{v^2}(m_3-m_1).
    \label{CP}
\end{equation}
 For the hierarchical SM neutrino masses, $m_3-m_1 \sim 0.05 $ eV~\cite{10.1093/ptep/ptac097}. By plugging in this maximum CP violation in the expression for baryon asymmetry (Eq.~\ref{BAU}),  one gets a lower bound $M_1\gtrsim 10^9$ GeV. On the other hand, $\epsilon_{max}$ can atmost be 1, which sets an upper bound $M_1\lesssim 10^{15}$ GeV.
 In presence of this CP violation, decays of the RHNs will produce the lepton asymmetry which then would get converted into the baryon asymmetry via Electroweak Sphalerons.

\subsubsection{Non-Thermal production of RHNs and BAU}
Within the framework of Type-I seesaw baryon asymmetry (via lepton asymmetry) can be achieved if some initial RHN abundance is present in the universe. In our scenario, this arises naturally when the RHNs are coupled to $\phi$ that undergoes a FOPT in the early universe. The RHNs are then produced from the decay of the bubble walls of $\phi$. Just like the DM case, we consider a Yukawa coupling $y_\phi$ between the scalar $\phi$ and RHNs $N$ along with the Seesaw lagrangian 
\begin{equation}
    -\mathcal{L} \supset \lambda_N \bar{L}(i\sigma_2)H^\dagger N + \frac{1}{2}M_N\bar{N}^cN + y_\phi \phi \bar{N}^cN + h.c.
    \label{51}
\end{equation}
The Yukawa coupling $y_\phi$ is responsible for the RHN production from the bubble wall decays. The goal is to reproduce the observed baryon asymmetry of the Universe 
\begin{equation}
    \mathcal{Y}_B \equiv \frac{n_B-n_{\bar{B}}}{s} = (8.69\pm0.22)\times 10^{-11},
    \label{52}
\end{equation}
where $n_B$, $n_{\bar{B}}$ and s are the number densities of baryons, antibaryons and entropy at present time. A convenient and simple parameterization of the baryon asymmetry produced in our leptogenesis scenario is~\cite{cataldi2024leptogenesisbubblecollisions}
\begin{equation}
    \mathcal{Y}_B = \mathcal{Y}_N\epsilon_{CP}c_{sph}\kappa_{wash}.
    \label{BAU}
\end{equation}
Here $\mathcal{Y}_N$ is the RHN yield produced from bubble collisions, $c_{sph}$ = 28/79 is the
sphaleron conversion factor~\cite{PhysRevD.42.3344}, $\epsilon_{CP}$ is the CP violation coming from the decays of RHN to SM given by Eq.~(\ref{CP}) and $\kappa_{wash}$ is efficiency factor parameterizing the washout effects from the various processes during leptogenesis. 
\noindent For non-thermal production, other washout effects are not present due to lack an interacting thermal plasma which otherwise would be present in any thermal leptogenesis scenario. For our purposes we work in the limit where wash out processes are negligible, $\kappa_{wash} = 1$. 

\subsection{Formation of Baryon Asymmetry and its GW probe}
\noindent Following last section, we find the generated Baryon asymmetry from bubble collision, given by Eq.~(\ref{BAU}) is
\begin{equation}
    \mathcal{Y}_B = \epsilon_{CP}c_{sph}\mathcal{Y}_N.
    \label{sixtyfour}
\end{equation}
For runaway phase transition with released vaccum energy of the form $\Delta V = \lambda v_\phi^4$, we have $\mathcal{Y}_N$ similar to Eq.~(\ref{46}) and the CP violation parameter $\epsilon_{CP}$ is given in Eq.~(\ref{CP}). With this, we first check whether for some values of the RHN mass $M_1$ and coupling $y_\phi$ we can get sufficient baryon asymmetry required by Eq.~(\ref{52}).

\begin{figure}[H]
    \centering
    \includegraphics[width=0.45\linewidth]{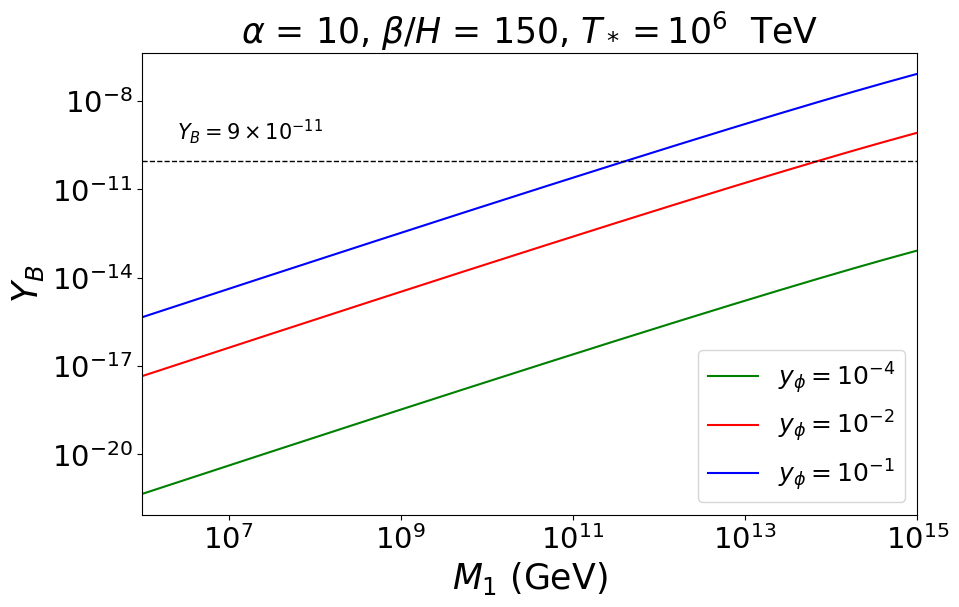}
    \includegraphics[width = 0.45\linewidth]{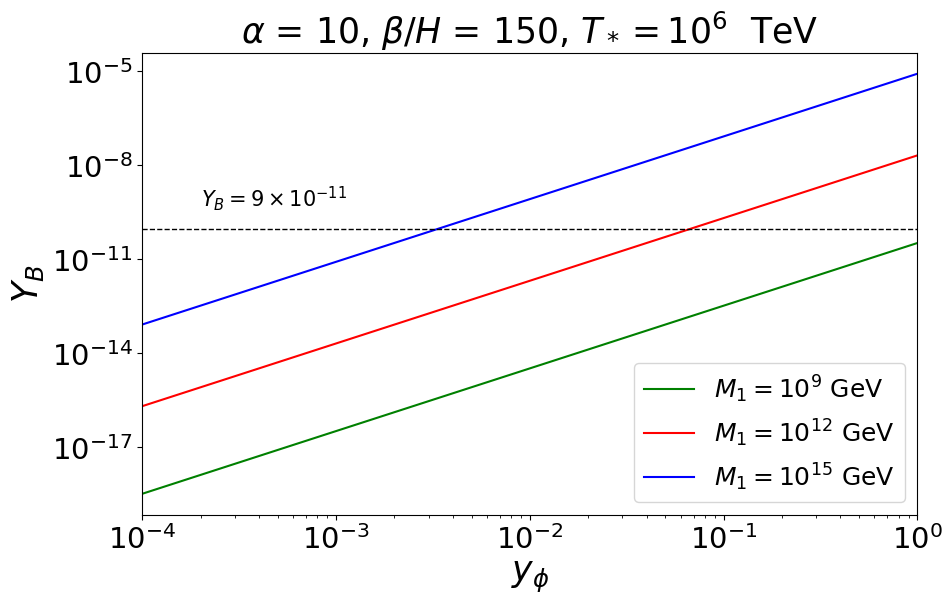}
    \caption{\it BAU for phase transition parameters \text{BP3}. \textbf{Left panel:} Baryon asymmetry vs RHN mass : plotted for three different benchmark points of the coupling strength. \textbf{Right panel:} Baryon asymmetry relic vs RHN coupling : again plotted for three different benchmark points of the RHN mass. In both cases $\lambda = \mathcal{O}(1)$ is taken. }
    \label{fig:6}
\end{figure}

\begin{figure}[H]
    \centering
    \includegraphics[width=0.7\linewidth]{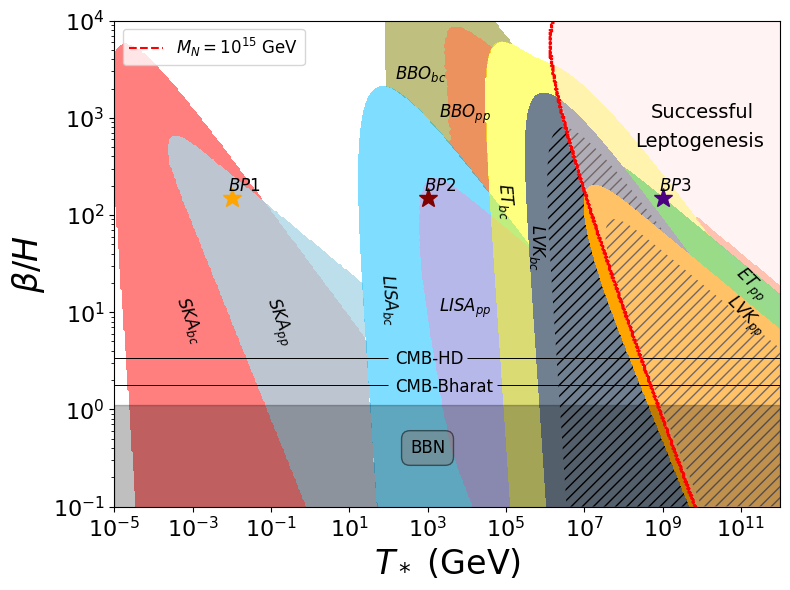}
    \caption{\it Regions with SNR $> 10$ plotted for different GW detectors for bubble collision and particle production mechanisms respectively for $\alpha = 10$, $y_\phi = 0.9$. On the same plane, three benchmark points BP1,BP2 and BP3 are plotted along with the region of correct observed baryon yield, $Y_B = 10^{-10}$ that can be achieved with different RHN masses. Here $\lambda \sim \mathcal{O}(1)$ is taken. Hatched regions are ruled out from Ligo-Virgo-Kagra (LVK) $O3$ observations. }
    \label{fig:9}
\end{figure}

\noindent In Fig~\ref{fig:6}, we see, for high phase transition temperature $(T_*)$, the required baryon asymmetry can be produced for RHN masses $M_1 \gtrsim 10^9$ GeV and couplings $y_\phi\gtrsim 10^{-2}$. A detailed parameter scan is left for Fig.\ref{fig:9} In general there are three Lagrangian parameters (see Eq.~\ref{51}) that can affect the BAU : $\lambda_N,\ y_\phi,\ M_1$ along with the three P.T. parameters : $\alpha,\ \beta/H,\ T_*$. Fixing the CP violation to be maximal fixes $\lambda_N$ (see Eq.~\ref{CP1}). This leaves us with five free parameters and calls for a detailed parameter scan. 

\begin{figure}[H]
    \centering
    \includegraphics[width=0.45\linewidth]{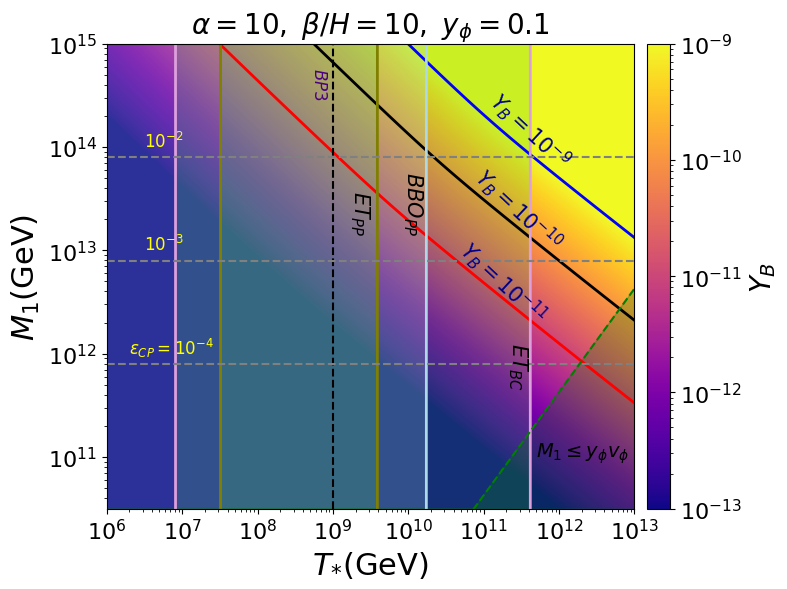}
    \includegraphics[width=0.45\linewidth]{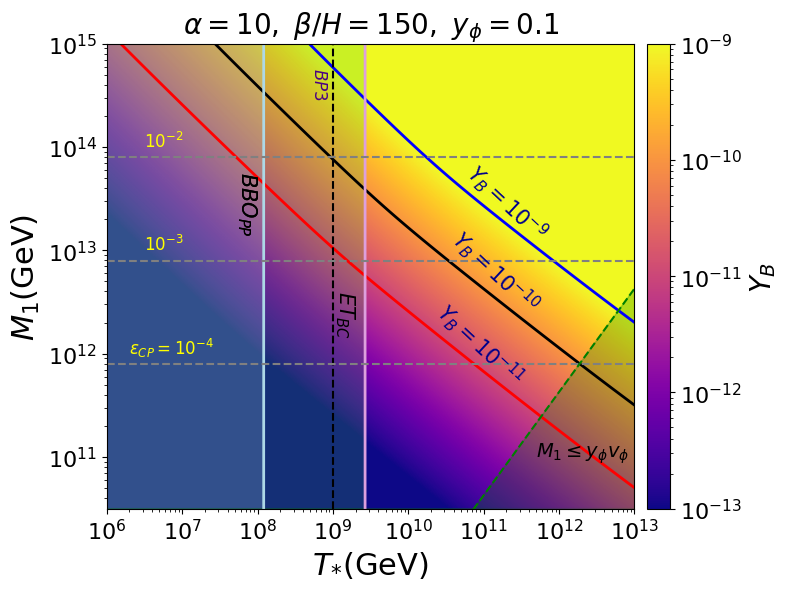}
    \caption{\it BAU yield plotted as a function of Phase transition temperature and the RHN mass for two benchmark points of $\beta/H$ with $y_\phi = 0.1$ and $\alpha = 10$. The shaded green region is where $\phi$ vev becomes comparable to RHN mass $M_1$. The vertical lines are for SNR = 10 of GW signals for different detectors (see the label). Left side of the vertical lines are regions with SNR $>10$. On the same plane, the line corresponding to BP3 is also shown. }
    \label{fig:8}
\end{figure}

\begin{figure}[H]
    \centering
    \includegraphics[width=0.47\linewidth]{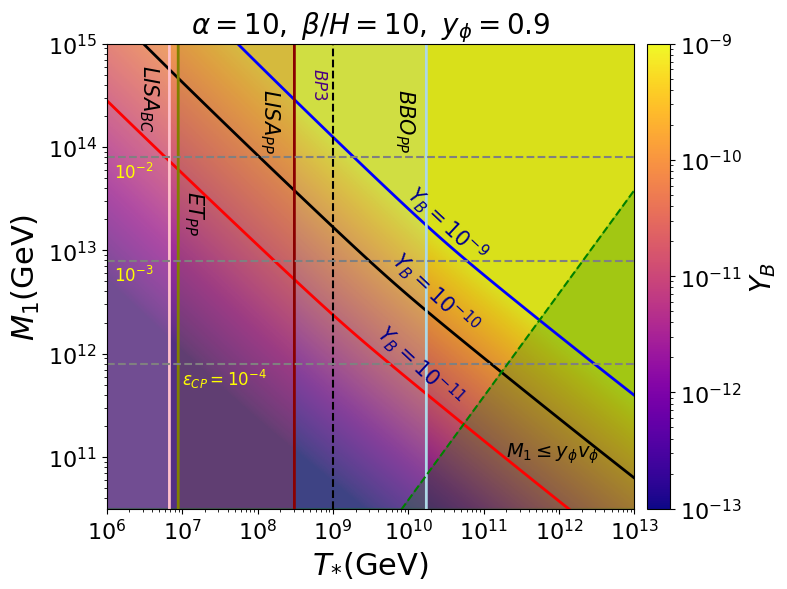}
    \includegraphics[width=0.47\linewidth]{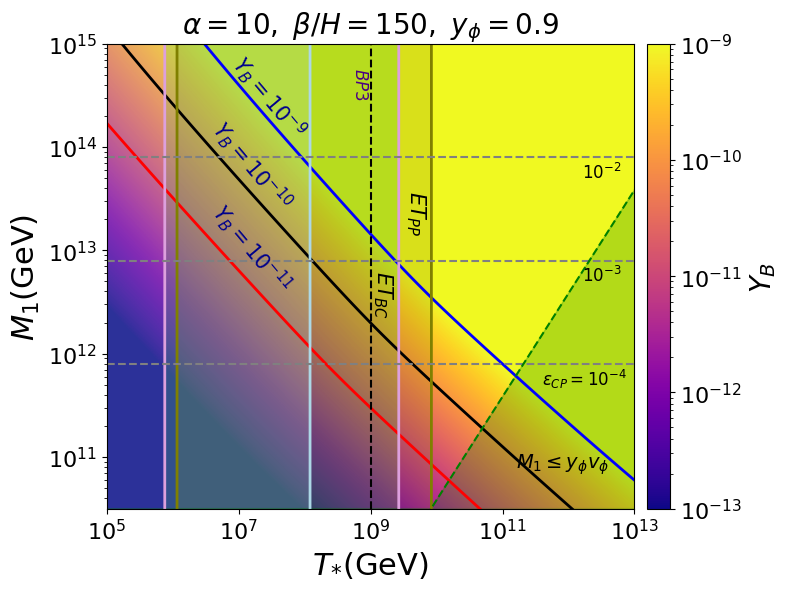}
    \caption{\it  BAU yield plotted as a function of Phase transition temperature and the RHN mass for two benchmark points of $\beta/H$ with $y_\phi = 0.9$ and $\alpha = 10$. The shaded green region is where $\phi$ vev becomes comparable to RHN mass $M_1$. The vertical lines are for SNR = 10 of GW signals for different detectors (see the label). Left side of the vertical lines are regions with SNR $>10$. Again the line corresponding to BP3 is shown on this plane. }
    \label{fig:nin3}
\end{figure}

Figures~\ref{fig:8} and \ref{fig:nin3} show that the observed baryon asymmetry can be reproduced in this model over a broad region of parameter space. For instance, for $\alpha = 10$, $\beta/H = 10$, and $y_\phi = 0.1$, successful baryogenesis is obtained for $T_* \gtrsim 10^{9}\ \mathrm{GeV}$ with RHN masses $M_1 \in [10^{12},10^{15}]~\mathrm{GeV}$. Portions of this region are also accessible to upcoming GW detectors: BBO and ET can probe the RHN-production GW signal, while ET is additionally sensitive to GWs from bubble collisions. This will allow for a inter-detector distinguishability scenario as mentioned in sec~\ref{III}, where receiving a bubble collision GW signal in ET along with a unique RHN production GW signal in BBO would conclusively prove the validity of particle production formalism.

For a faster transition with $\beta/H = 150$, the correct baryon asymmetry can be achieved at even lower temperatures, $T_* \in [10^{7},10^{12}]~\mathrm{GeV}$ for the same $M_1$ range, although the corresponding GW detectability is reduced. Increasing the coupling to $y_\phi = 0.9$ allows successful baryogenesis down to $T_* \simeq 10^{6}~\mathrm{GeV}$. In this case, parts of the parameter space remain testable with ET, BBO, and LISA, with all three detectors exhibiting overlapping regions where $\mathrm{SNR}>10$. This is clearly seen in Fig.~\ref{fig:nin3} where LISA and ET both probes the RHN production GW spectrum for a portion of the parameter space and hence again can be used for a inter-detector sensing.

The green shaded regions on the right-hand side indicate where the scalar vev becomes comparable to the RHN mass, rendering the baryon-yield calculation unreliable. In this regime the RHNs obtain their mass predominantly from the phase transition, and, as noted earlier, the particle-production formalism break down \cite{shakya2025aspectsparticleproductionbubble}.

Figures~\ref{fig:8} and \ref{fig:nin3} then indicate that successful baryogenesis in our framework will be probed via detectable GW signals in certain regions of parameter space, with the possibility of GW features of characteristic of the RHN production from the bubbles. To examine this interplay more closely, we next present the regions yielding the correct baryon asymmetry together with the corresponding GW SNRs in the $\beta/H$–$T_*$ plane.  

We show in Fig.~\ref{fig:9} the detection thresholds for the GW signals associated with the leptogenesis scenario described above. On the same plane, assuming maximal CP violation, we show the region of parameter space that yields sufficient baryon asymmetry for RHN masses in the allowed range $M_1\in[10^9,10^{15}]$ GeV. Although part of the parameter space is already excluded by LVK searches (orange and gray hatched regions), future detectors such as BBO, ET and future LVK observations retain significant discovery potential. For example, for a very heavy RHN mass $M_1 = 10^{15}$ GeV and a fast phase transition, $\beta/H \sim 10^3$, one expects a bubble-collision signal within the ET and future LVK sensitivity bands together with a RHN-production GW signal in the BBO frequency range at temperatures $T_* \sim 10^7$ GeV. As the RHN-production contribution exhibits a characteristic peaked spectrum, with a low-frequency rise (approximately $f^1$), this again enables one to conclusively probe the RHN-production mechanism and leptogenesis with a inter-detector probing. 

The RHN mass is bounded from below by the Davidson–Ibarra bound (see Eq.~\ref{CP}). It also cannot exceed the upped bound of $10^{15}$ GeV coming from maximal CP violation. As a result, viable leptogenesis in this framework requires relatively high phase transition temperatures, $T_*\gtrsim10^9$ GeV. A sizable portion of the parameter space therefore is already constrained by LVK observations. Nonetheless, a region with sufficiently rapid transitions remains viable. By inspection of Fig.~\ref{fig:9}, the surviving band is approximately characterized by \begin{equation*}
    \beta/H\gtrsim100 ,\quad T_*\gtrsim 10^8 \ \rm{GeV},\quad 10^9\ \rm{GeV}\lesssim M_1\lesssim 10^{15}\ \rm{GeV},
\end{equation*} which lies within the projected sensitivities of BBO and ET while remaining outside the LVK-excluded domain. In particular, the benchmark point BP3 satisfies all these conditions and offers a concrete target for upcoming GW searches. Furthermore, at very high P.T. temperatures $T_*\sim 10^{11}$ GeV, there exists a region of parameter space with $\beta/H\sim 20-30$ where only the GW signal from RHN production is detectable by BBO and ET. The parameter space lies outside the region excluded by LVK observations and in this regime, GW signals from bubble collisions are entirely unobservable. Consequently, the detection of a GW signal in this region would constitute a clear smoking gun for new physics.


\section{AsDM via RHN decays produced from bubbles}
\label{V}
\noindent In the previous section, we discussed how stable RHNs can constitute viable dark matter candidates. We also showed that the same RHNs, when rendered unstable or meta-stable, can generate the baryon asymmetry via leptogenesis while simultaneously accounting for the masses of the active SM neutrinos. In both cases, the distinctive GW signals arising from the RHN production mechanism, together with the well-known bubble-collision signals, remain detectable in several current and upcoming GW detectors.

\medskip

Besides these, current observations show that the DM and Baryon relics are very close today, with $\Omega_{DM}/\Omega_B \sim 5$ 
which is roughly within the same order of magnitude.\footnote{Note, as compared to this, for the Dark Energy relic ($\Omega_\Lambda$) observed today one has $\Omega_\Lambda/\Omega_b \gtrsim \mathcal{O}(10)$. } 
This suggests the possibility of a common origin for both relics. One class of scenarios that naturally relates the baryon and dark matter abundances is Asymmetric Dark Matter (AsDM)~\cite{Nussinov:1985xr,Kaplan:1991ah}. In such models, the dark matter particle possesses a distinct antiparticle, and comparable asymmetries in the visible and dark sectors are generated by similar microphysics, for instance through the decay of same heavy particle. 

\medskip

Building on this idea, several works have explored asymmetric dark matter arising from RHN decays during leptogenesis : for $n_{DM}\sim n_{b}$  scenarios, where an asymmetry is first produced in one sector and is then transferred to the other sector at later times via some portal interactions~\cite{Cosme_2005,Gu_2009} and for a two sector leptogenesis \cite{Falkowski:2011xh} where asymmetry is generated in both visible and dark sectors simultaneously. %

\medskip

In what follows, we consider a two-sector leptogenesis framework similar to Ref.~\cite{Falkowski:2011xh}. However, unlike the conventional setup, the initial RHN abundance in our scenario is generated non-thermally through production from runaway bubble collisions during a first-order phase transition in the early universe (see Fig.~\ref{fig:diag}). As we show below, this leads to a testable realization of co-genesis, since the associated RHN-production mechanism imprints a distinctive GW signature that can be probed in upcoming experiments.

\subsection{Leptogenesis and Asymmetric Dark Matter}
In this setup the DM field $\chi$ resides in a hidden Dark Sector (DS)
indirectly connected to the SM via Yukawa interactions with heavy Majorana neutrinos, $N$. The SM leptons and the DM particle are charged under an approximate lepton number, which is broken by the Majorana masses of $N$. The Yukawa couplings can be complex, leading to CP violation in the decays of $N$. The generation of the DM abundance adheres to the following steps,
\begin{itemize}
    \item A population of the lightest Majorana neutrino, $N_1$, is generated in the early
universe with a mass $M_{1}$ from a first order phase transition.
    \item At temperatures below $M_{1}$, these neutrinos decay out of equilibrium to both sectors. The CP-violating decays lead to a lepton number asymmetry in both the SM and hidden sector.
    \item As the universe cools well below $M_{1}$, the washout of lepton asymmetry, and its transfer between the 2 sectors, becomes inefficient and the asymmetries are frozen-in. The asymptotic asymmetry can, in general, be different in the two sectors due to different branching fractions and/or washout effects.
    \item the symmetric component of the DM number density is annihilated away in the hidden sector. The relic abundance of DM is set by the remaining asymmetric component much like SM baryogenesis scenario. 
\end{itemize}
To realize these, we consider the lagrangian in Eq.~(\ref{51}) with addition of a new term involving the Yukawa interaction between the Fermionic DM state $\chi$, the RHN $N$ and the scalar driving the phase transition

\medskip

\begin{equation}
    -\mathcal{L} \supset \lambda_N \bar{L}(i\sigma_2)H^\dagger N + \frac{1}{2}M_N\bar{N}^cN + y_\phi \phi \bar{N}^cN + y_\chi N\chi\phi + h.c.
    \label{sixtyeight}
\end{equation}

\medskip

Here $y_\chi$ is the interaction strength between the RHN and the DM. We take $\chi$ such that $\chi\phi$ is a gauge singlet under the symmetry that is broken by $\phi$ vev. Note that, at the time of bubble collision, some amount of $\chi$ particles would also be generated from $y_\chi N\chi\phi$ term. Also after $\phi$ gets a vev, there would be mixing between $N$ and $\chi$ particles due to the same term in the Lagrangian. To ensure bubble collisions predominantly produce RHNs and the mixing between $\chi$ and $N$ is negligible, we will take $y_\phi \sim \mathcal{O}(1) \gg y_\chi$.

\medskip

In order to generate an asymmetry in leptons and in DM, there must be CP-violation in the decays of $N$. We take the hierarchal approximation~\cite{Falkowski:2011xh}, $ M_1 \ll M_2$ and consider only the asymmetry generated by the decays of $N_1$ :
\begin{equation}
    \epsilon_\chi = \frac{\Gamma(N_1\rightarrow \chi\phi) - \Gamma(N_1\rightarrow \bar{\chi}\phi^\dagger)}{\Gamma_{tot}} \ ;\qquad \epsilon_l = \frac{\Gamma(N_1\rightarrow \bar{l}h) - \Gamma(N_1\rightarrow lh^\dagger)}{\Gamma_{tot}},
\end{equation}
where the total decay width for the RHN is $\Gamma_{tot} = \Gamma_1+\Gamma_\chi$, with $\Gamma_1$ defined in Eq.~(\ref{56}) and $\Gamma_\chi$ given as~\cite{Datta:2025vyu}
\begin{equation}
    \Gamma_\chi = \frac{|y_\chi y_\chi^\dagger |_{11}}{16\pi}M_1.
    \label{seventy seven}
\end{equation}
Notice the extra 1/2 factor in Eq.~(\ref{seventy seven}) as compared to Eq.~(\ref{56}) is there because  this decay has singlets and not doublets in the final state.  

\medskip

The asymmetries are sourced by the interference between tree and loop diagrams of the $N_1$ decay. The CP violation arising from RHN decaying into SM states has already been discussed and is given in Eq.~(\ref{CP1},\ref{CP}). Note here, due to additional coupling of RHNs to the DS, there are additional loops contributing to the CP violation.

\medskip

\begin{figure}[H]
\centering
\includegraphics[width = 0.8\linewidth]{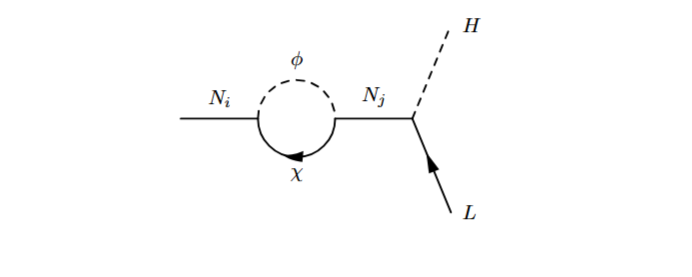}
\caption{\it Self energy contribution to $N_1 \rightarrow LH$ changes due to DS loop}
\end{figure}

\medskip

\noindent Hence the maximum CP violation and the Davidson-Ibarra bound gets slightly modified~\cite{Falkowski:2011xh} 
\begin{equation}
    \epsilon_L \leq \frac{3}{16\pi}\frac{M_1}{v^2}\Delta m_\nu \cdot C ,
    \label{69}
\end{equation}
where C is a constant which depends on $M_1$ and the yukawa matrices and reduces to $C = 1$ in the limit when RHN predominantly decays into SM. The CP violation in the DS depends on the coupling of RHN to $\chi$ particles and RHN masses and is given as 
\begin{equation}
    \epsilon_\chi \simeq \frac{1}{16\pi}\frac{M_1}{M_2}| y_\chi y_\chi^\dagger|_{22}.
\end{equation}

\medskip

\noindent As the RHN $N_1$ decays, these asymmetries leads to asymmetry in both sectors. We define the branching ratios of the RHN $N_1$ to both sectors as

\begin{equation}
    \text{Br}_L = \frac{\Gamma_1}{\Gamma_1+\Gamma_\chi}\ ; \quad \text{Br}_\chi = \frac{\Gamma_\chi}{\Gamma_1+\Gamma_\chi}.
\end{equation}
Then the produced baryon asymmetry and DM yield can be parametrized
as\cite{Datta:2025vyu}
\begin{equation}
    \mathcal{Y}_B = \frac{28}{79}\epsilon_L\eta_L\text{Br}_L \mathcal{Y}_N^i \ ; \quad \mathcal{Y}_{DM} = \epsilon_\chi \eta_\chi \text{Br}_\chi \mathcal{Y}_N^i,
    \label{73}
\end{equation}
where $\mathcal{Y}_N^i$ is the initial yield of the RHNs produced from the bubble collision, given by Eq.~(\ref{46}) and $\eta_\chi$ , $\eta_L$ are the washout efficiencies of the DS and SM sector asymmetries respectively. The relevant processes contributing to the washout are 

\medskip

\begin{enumerate}
    \item \textbf{Inverse Decay :} The inverse decays $LH\rightarrow N_1$ and $\chi\phi\rightarrow N_1$ can wash out the produced asymmetries. These processes will be boltzmann suppressed.
    \item \textbf{$2\leftrightarrow2$ Scattering :} Scattering processes such as $LL\leftrightarrow HH$, $LH\leftrightarrow \chi\phi$ etc can contribute to mixing of the asymmetries in both sectors and alter final asymmetries. 
\end{enumerate}

\noindent In the narrow width approximation with $\Gamma_{tot}\ll M_1$ and $\Gamma_{tot}^2/M_1H_1 \ll 1$, the inverse decays play major role for washout, whereas in the large transfer regime with $\Gamma_{tot} \simeq M_1$ or $\Gamma_{tot}^2/M_1H_1 \gtrsim 1$ the $2\leftrightarrow2$ scatterings dominate.

\medskip

We will be working in the narrow width regime where one can ignore the scattering effects and the final asymmetries become uncorrelated. Furthermore, since the inverse decay processes are Boltzmann suppressed $\Gamma_{inv} \sim \exp{-M_1/T_*}$, we can safely ignore any washout effects and take $\eta_{L,\chi} = 1$. 

\subsection{AsDM formation}
The DM yield can be related to the DM mass $m_{DM}$ as below \cite{Datta:2025vyu}
\begin{equation}
    \mathcal{Y}_{DM} \simeq 4.37\times 10^{-10}\left(\frac{\Omega_{DM}h^2}{0.12}\right)\left(\frac{\text{GeV}}{m_{DM}}\right),
\end{equation}
where $\Omega_{DM}$ is the DM abundance observed today.  Since $\chi$ here is our DM candidate, $m_{DM} = m_\chi$ and we use these two notations interchangeably throughout the draft. Similarly the observed baryon asymmetry today is given by Eq.~(\ref{52}). 

\medskip

Assuming a hierarchy between the RHN masses 
$M_1:M_2:M_3 = 1 : 10 : 100$,  we get $\epsilon_\chi \lesssim 10^{-2}$~\cite{Falkowski:2011xh}. We also assume maximal CP violation for $\epsilon_L$ given in Eq.~(\ref{69}). This leaves us with a total of three Lagrangian parameters : $M_1,\ y_\phi,\ y_\chi$ and three P.T. parameters $\alpha,\ \beta/H,\ T_*$ that affects the Baryon and DM relic. We trade $y_\chi$ for the Branching ratio $\rm{Br}_\chi$ and look for the parameter space for AsDM which satisfy the leptogenesis condition as well as active neutrino masses.\\

\begin{figure}[H]
    \centering
    \includegraphics[width=0.48\linewidth]{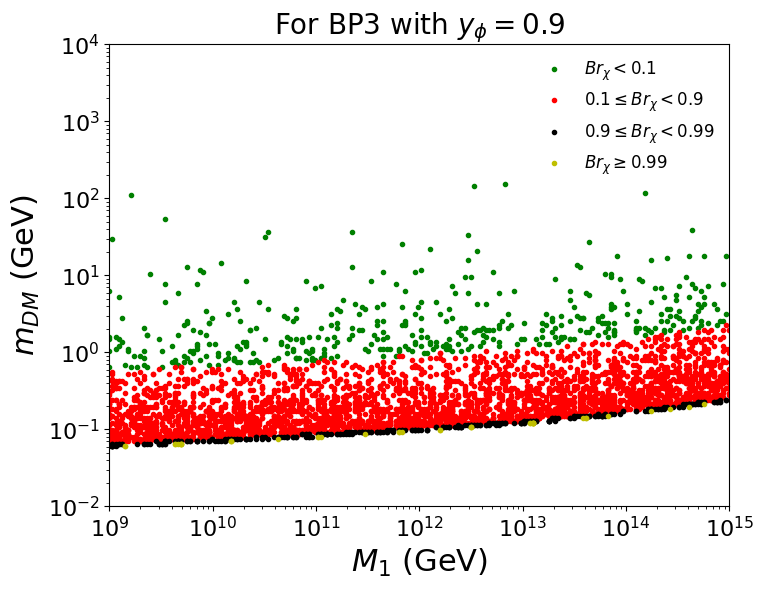}
    \includegraphics[width=0.48\linewidth]{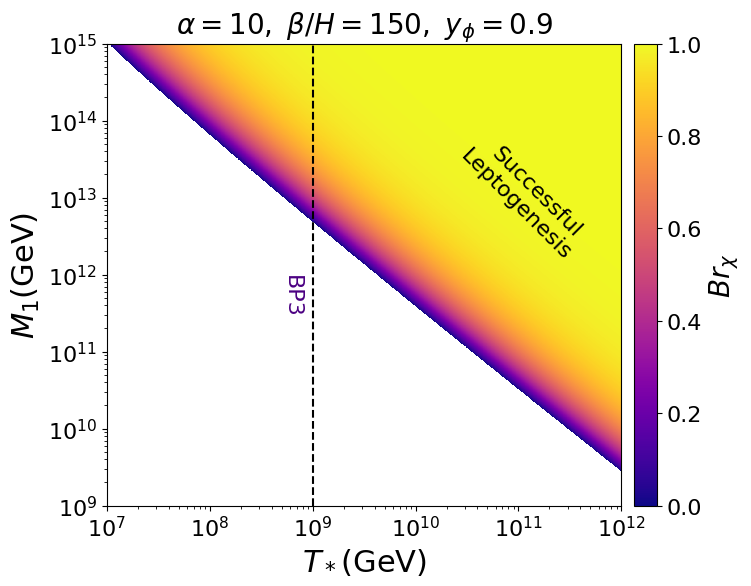}
    \caption{\it Parameter space for two-sector non-thermal leptogenesis from a first-order phase transition, shown for $\beta/H = 150,\ \alpha = 10$ with$\ y_\phi = 0.9$. \textbf{(a) Left :} Points yielding the observed dark matter abundance today, plotted as a function of the RHN mass $M_1$ for different values of the branching ratio $\text{Br}_\chi$. We have taken $T_* = 10^{9}$ GeV (BP3) and $\epsilon_\chi = 10^{-2}$ corresponding to the maximal CP asymmetry attainable for the chosen RHN mass hierarchy. \textbf{(b) Right:} Parameter space consistent with successful leptogenesis. The maximal value of $\epsilon_L$ is used for this panel, and the required $\text{Br}_\chi$ leading to the observed baryon asymmetry is indicated. The benchmark line corresponding to BP3 is shown for reference.}
    \label{fig:twelve}
\end{figure}

Figure~\ref{fig:twelve}(a) shows that for BP3 we can successfully obtain non-thermal asymmetric dark matter, with a lower bound on the DM mass of approximately $m_{DM}\gtrsim 100$ MeV, which arises from the very high phase transition temperature $T_* = 10^9$ GeV that sets the scale of the initial RHN abundance. We find that across the entire allowed range of RHN masses, asymmetric dark matter can be realized for a broad range of branching ratios $\rm{Br}_\chi$. On the other hand, Fig.~\ref{fig:twelve}(b) displays the region consistent with successful leptogenesis together with the value of the branching ratio $\rm{Br}_\chi$ required to reproduce the observed baryon yield. The BP3 line shows that leptogenesis becomes efficient for RHN masses $M_1\gtrsim10^{13}$ GeV. Taken together, these observations indicate that BP3 admits a viable window in which both the baryon asymmetry and the dark matter relic abundance can be simultaneously accommodated, provided the required branching ratio remains compatible. This motivates a more systematic exploration of the parameter space for co-genesis, which we undertake next.

\subsection{GW probe of AsDM}
\begin{figure}[H]
    \centering
    \includegraphics[width=0.49\linewidth]{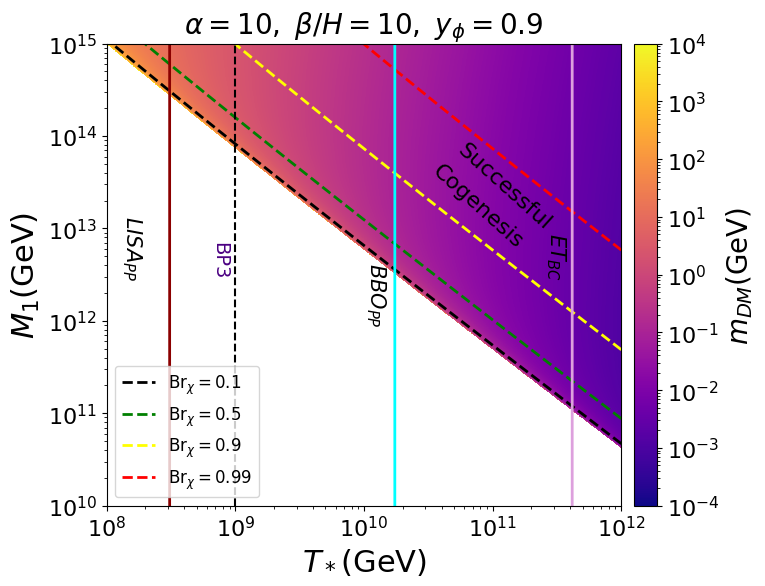}
    \includegraphics[width=0.49\linewidth]{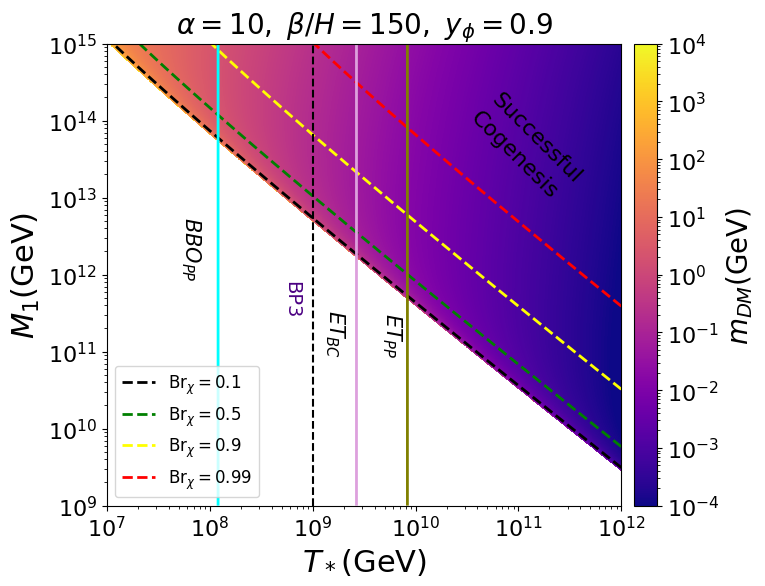}
    \caption{\it Parameter scan for successful co-genesis as a function of the phase–transition temperature $(T_*)$ and RHN mass $(M_1)$ for two benchmark choices with $y_\phi = 0.9$ and $\alpha = 10$. \textbf{Left :} Relatively slow transition with $\beta/H = 10$ \textbf{Right :} Faster transition with $\beta/H = 150$ (BP3). On each plane, we overlay the lines corresponding to SNR = 10 for various GW detectors (see label), including contributions from both particle-production and bubble-collision mechanisms. The region left to each SNR curve corresponds to SNR $> 10$. For parameter points yielding the correct baryon asymmetry, the predicted dark-matter mass is indicated by the color bar. We assume maximal CP violation in both sectors, with $\epsilon_\chi \sim10^{-2}$ and $\epsilon_L$ given in Eq.~(\ref{69}).  $\lambda\sim\mathcal{O}(1)$ taken for this plot.}
    \label{fig:13}
\end{figure}

Fig.~\ref{fig:13} shows the region of parameter space that yields successful co-genesis of AsDM and the baryon asymmetry. For both slow and relatively fast phase transitions—corresponding to low and high values of $\beta/H$—we find that a broad range of RHN masses, $M_1 \in [10^{10},10^{15}]$ GeV, and phase transition temperatures, $T_* \in [10^8,10^{12}]$ GeV, satisfies the co-genesis requirements. The resulting dark matter mass may vary widely, with a lower bound of $m_{\rm DM} \gtrsim 0.1$ MeV and an upper bound of $m_{\rm DM} \lesssim 10$ TeV. Consequently, even though the scale of the phase transition and leptogenesis is extremely high, the DM candidate need not be ultra-heavy. 

The associated GW signals produced during the FOPT provide complementary detection prospects. For $\beta/H = 10$, LISA and BBO can probe the particle-production component of the spectrum with $m_{\rm DM}$ in the GeV–TeV range, while the bubble-collision signal is accessible to ET. Hence observing the bubble collision GWs in ET and a corresponding RHN production GW spectrum for the same P.T. parameters in LISA or BBO would strongly point towards a FOPT sourcing baryon asymmetry and DM. For a relatively fast transition with $\beta/H = 150$, the GW signals associated with $m_{\rm DM} \gtrsim 1$ GeV become observable in ET for both bubble-collision and RHN-production channels. In contrast, BBO is sensitive only to the RHN-production component for comparatively heavier dark-matter masses, $m_{\rm DM} \gtrsim 100\ \mathrm{GeV}$. Thus, a detection of the bubble-collision signal in ET, followed by an independent detection of the RHN-production signal in BBO, would provide a clear inter-detector consistency test of the underlying co-genesis mechanism. 

\medskip

Overall, this study demonstrates that asymmetric DM and baryon co-genesis can naturally arise from a high-scale FOPT, and that the corresponding GW signals, especially those from the novel RHN production mechanism offers promising avenues for detection in upcoming interferometers.

\medskip

\section{Viable UV-complete Particle Theory: Multi-Majoron Model}
\label{VI}

From the observations of various neutrino oscillation experiments, the mass square splittings of the left-handed neutrino masses have been measured as $\Delta m_{21}^2 \simeq 7.5 \times 10^{-5} \mathrm{eV}^2 $ and $\Delta m_{3\ell}^2 \simeq 2.5 \times 10^{-3} \mathrm{eV}^2 $ \cite{Esteban:2024eli}. Beside this, the cosmological bound for the sum of left-handed neutrino masses is approaching $0.1$ eV \cite{Palanque-Delabrouille:2019iyz, eBOSS:2020yzd}. 
Both these indicate the left-handed neutrino mass spectrum is likely to be hierarchical. 

\medskip

\noindent The right-handed neutrinos involved in the type I seesaw mechanism are also expected to have a strong hierarchical mass spectrum in several extensions in general, and also motivated from the explaining the puzzle of hierarchial fermion masses in the SM \cite{Froggatt:1978nt,Barr:1979xt,Barbieri:1983uq,Balakrishna:1987qd,Babu:1988fn,Babu:2004th,Babu:2009fd}, see Refs.~\cite{Babu:1999me,Yoshioka:2000tve,Chao:2012re,Altmannshofer:2014qha,Higuchi:2014cua,Huitu:2017ukq,Okada:2019fgm,Hernandez:2021iss,Bonilla:2023wok,FernandezNavarro:2024hnv,Arbelaez:2024rbm,Jana:2024icm} for attempts to bring the seesaw explanation of neutrino mass and fermion mass hierarchy under one umbrella \footnote{Models with sequential dominance \cite{King:2013eh,Fu:2023nrn} or vanilla leptogenesis \cite{Buchmuller:2004nz,Davidson:2008bu} also predict such hierarchical RHN masses.}.

\medskip

\noindent In this section, as an illustrative example only, we consider a minimal extension of the SM explaining the hierarchical right-handed neutrino masses in what is known as multi-Majoron models \cite{Bamert:1994hb}, where different right-handed neutrinos $N_i$ couple to complex scalar Majoron fields $\phi_i$ through Yukawa interaction which we describe in detail below. The model we consider is based on a global ${\rm U}(1)_{B-L}\times\rm{U}(1)_N$ extension of the SM gauge group, $SU(3)_c\times SU(2)_L\times U(1)_Y$ \cite{Fu:2025dlp, DiBari:2023mwu}. 

\begin{table}[H] \centering
\begin{tabular}{|c||c|c|c|c|c|}
\hline
& $SU(3)_c$&$SU(2)_L$& $U(1)_Y$& $U(1)_{B-L}$& $U(1)_N$ \\ \hline\hline
$q_L^i$& {\bf{3}}&{\bf 2}&+1/6&+1/3&0\\
$u_R^i$& {\bf{3}}&{\bf 1}&+2/3&+1/3&0\\
$d_R^i$& {\bf{3}}&{\bf 1}&$-1/3$&+1/3&0\\ \hline
$l_L^i$& {\bf{1}}&{\bf 2}&+1/6&$-1$&0\\
$e_R^i$& {\bf{1}}&{\bf 1}&$-1$&$-1$&0\\
$N_3$& {\bf{1}}&{\bf 1}&0&$-1$& $0$\\
$N_{1,2}$& {\bf{1}}&{\bf 1}&0&$0$& $-1$\\ \hline
$H$& {\bf{1}}&{\bf 2}&$-1/2$&0&0\\
$\Phi_1$& {\bf{1}}&{\bf 1}&0&+2&0\\
$\Phi_2$& {\bf{1}}&{\bf 1}&0&0&+2\\ \hline
\end{tabular}
\caption{\it Particle content of the two-Majoron model. }
\label{table_matcon}
\end{table}
\noindent Three generations of RHNs  $N_i$ ($i = 1,2,3$) are introduced along with two additional complex scalar fields $\Phi_1,\Phi_2$, charged under the global $U(1)_{B-L}$ and $U(1)_N$ respectively. These scalars acquiring vev spontaneously breaks the $U(1)_{B-L}\times U(1)_N$ symmetry, which contribute to the masses of the RHNs. The particle content of the model is listed in Table~\ref{table_matcon}. The additional Yukawa terms involving the RHNs are given by ~\cite{DiBari:2023mwu}: 
\begin{align}
\mathcal{L}_Y
&\supset -\lambda_N^{ij}\bar{N}_i H^\dag l_L^j-\frac{1}{2}y_{3}\Phi_1\bar{N}^{c}_3N_3 -\frac{1}{2}y_{1}\Phi_2\bar{N}^{c}_{1}N_{1} -\frac{1}{2}y_{2}\Phi_2\bar{N}^{c}_2N_{2} +\text{h.c.} ,
\label{seventyfive}
\end{align}
where the first term gives the Dirac neutrino mass after electroweak symmetry breaking, while the next terms generate the RHN Majorana masses $M_i$. We write the complex scalars as $\Phi_1 = \phi_1 e^{i\theta_1}/\sqrt{2}$ and $\Phi_2 = \phi_2 e^{i\theta_2}/\sqrt{2}$ and take the vev along the real axis, $\langle\Phi_1\rangle = v_1/\sqrt{2},\ \langle\Phi_2\rangle = v_2/\sqrt{2}$. We further assume the hierarchy $v_1\gg v_2$ such that the Majorana masses satisfy $M_3\gg M_1\approx M_2 \approx M$. Note this masses enter in the seesaw expression in Eq.~(\ref{fiftwo}) and give the SM neutrino masses : $m_{\nu_{1,2}} \sim v_h^2/M,\ m_{\nu_3} \sim v_h^2/M_3$, $v_h$ being the Higgs vev. Thus  Multi-Majoron model seesaw generates a hierarchy between the SM neutrino masses as well, which is different than standard single Majoron. Since $M \sim y_1 v_2/2,\ M_3\sim y_3 v_1/2$, the vevs $v_1,v_2$ of the $U(1)_N$ and $U(1)_{B-L}$ breaking scalars $\Phi_2,\Phi_1$ play important role in the seesaw mechanism which can further be probed via the novel GW signature.\\

\noindent Along with the usual quadratic and quartic terms, the symmetry $\rm U(1)_N \times \rm U(1)_{B-L}$ also allows a quartic mixing term in the tree level potential and can be written as
\begin{equation}
    V_0(\Phi_1,\Phi_2) = -\mu_1^2\Phi_1^\dagger\Phi_1 + \lambda_1(\Phi_1^\dagger\Phi_1)^2 -\mu_2^2\Phi_2^\dagger\Phi_2 + \lambda_2(\Phi_2^\dagger\Phi_2)^2 + \zeta(\Phi_1^\dagger\Phi_1)(\Phi_2^\dagger\Phi_2).
    \label{seventysix}
\end{equation}
At high temperatures both symmetries are restored. The $U(1)_{B-L}$ symmetry is broken at $T\sim v_1$, generating the mass $M_3$ for the heaviest RHN $N_3$. GWs produced during this transition are negligible~\cite{DiBari:2023mwu}. After $\Phi_1$ settles at its vev, the $\Phi_2$ phase transition occurs. Owing to the scalar mixing term, the zero-temperature effective potential for $\phi_2$ develops a cubic term in the mass-diagonal basis\footnote{See Eq.~(3.8) of Ref.~\cite{DiBari:2023mwu}.} : $\{\varphi_1,\varphi_2\}$, leading to a strongly first-order phase transition.\\

\noindent The finite temperature effective potential at one loop is given by
\begin{equation}
    V_{eff}^T(\varphi_2) \simeq V_0(\varphi_2)+ V_1^0(\varphi_2)+ V_1^T(\varphi_2),
\end{equation}
which for our case reduces to~\cite{DiBari:2023mwu}
\begin{equation}
    V_{eff}^T(\varphi_2) \approx \frac{1}{2}\tilde{M^2_T}\varphi_2^2-(AT+C)\varphi_2^3 + \frac{1}{4}\lambda_T\varphi_2^4.
    \label{seventyeigth}
\end{equation}
Here we have neglected $\mathcal{O}\left((v_2/{v_1})^2\right)$ or higher order terms. The different constants are
\begin{align*}
    C &= \frac{\zeta^2 v_2}{4\lambda_1}\ , \quad A = \frac{(3\lambda_2)^{3/2}}{12\pi}\ ,\quad \tilde{M_T^2} = 2D(T^2-T_0^2),\\
    \lambda_T &= \lambda_2 - \frac{M^4}{4\pi^2v_2^4}\log\left(\frac{a_F T^2}{e^{3/2}M^2}\right)+\frac{9\lambda_2^2}{16\pi^2}\log\left(\frac{a_B T^2}{e^{3/2}m_{\varphi_2}^2}\right),
\end{align*}
where $m_{\varphi_2}$ is the mass of the scalar $\varphi_2$ after $\phi_1$ and $\phi_2$ both acquires their VEVs 
\begin{equation*}
    m_{\varphi_2}^2 = -\mu_2^2 - 6Cv_2 + 3\lambda_2v_2^2\ ;
\end{equation*}
$a_F, a_B$ are constants arising due to the high-temperature expansion of the fermionic / bosonic thermal function and 
\begin{align*}
    2DT_0^2 = \lambda_2v_2^2 + \frac{M^4}{4\pi^2v_2^2} - \frac{3}{8\pi^2}\lambda_2^2v_2^2\ ; \quad D = \frac{\lambda_2}{8}+\frac{M^2}{12v_2^2} .
\end{align*}

Next, from this effective thermal potential we calculate the  Euclidean action $S_E$. Note for temperature dependent potentials the least action solution has $\mathcal{O}(3)$ symmetry and $S_E = S_3/T$ where $S_3$~\cite{Coleman:1977py},\cite{Linde:1981zj} is the $\mathcal{O}(3)$ symmetric Euclidean action
\begin{equation}
    S_3(\varphi_2,T) = \int d^3x \left[\frac{1}{2}(\bm{\nabla}\varphi_2)^2+V_{eff}^T(\varphi_2)\right] = 4\pi\int_0^\infty dr\ r^2\ \left[\frac{1}{2}\left(\frac{d\varphi_2}{dr}\right)^2+V_{eff}^T(\varphi_2)\right].
\end{equation}
The physical solution $\varphi_2$ that minimizes $S_3$ can be found by solving the EOM 
\begin{equation}
    \frac{d^2\varphi_2}{dr^2} + \frac{2}{r}\frac{d\varphi_2}{dr} = \frac{dV_{eff}^T}{dr}
    \label{eighty}
\end{equation}
 with the boundary conditions $\left(d\varphi_2/dr\right)_{r=0} = 0$ and $\varphi_2\left(r\rightarrow\infty\right) = 0$. It can be shown~\cite{Megevand:2016lpr} that at the phase transition temperature $T_*$ the Euclidean action $S_3$ has to satisfy
\begin{equation}
    \frac{S_3(T_*)}{T_*}-\frac{3}{2}\ln\left(\frac{S_3(T_*)/T_*}{2\pi}\right) = 4\ln\left(\frac{T_*}{H_*}\right) -4\ln\left[T_*\frac{S_3'(T_*)}{T_*}\right] + \ln(8\pi v_w^3).
    \label{eightyone}
\end{equation}
Here $H_* = H(T_*)$ is the Hubble parameter at the time of transition. Then Eq.~(\ref{eighty},\ref{eightyone}) together gives the value of the P.T. temperature $T_*$. The other P.T. parameters are calculated using the following relations 
\begin{equation}
    \frac{\beta}{H_*} \approx T_*\left.\frac{d\left(S_3/T\right)}{dT}\right|_{T = T_*}\ ;\qquad \alpha \equiv \frac{\xi(T_*)}{\rho(T_*)}.
\end{equation}
Here $\rho(T_*)$ is the total energy density of the plasma and $\xi(T_*)$ released latent heat during the phase transition,
\begin{equation}
    \xi(T_*) = -\Delta V_{eff}^{T_*}(\varphi_2)-T_*\Delta s(T_*) = -\Delta V_{eff}^{T_*}(\varphi_2) + T_*\left.\frac{\partial \Delta V_{eff}^{T_*}(\varphi_2)}{\partial T}\right|_{T_*},
    \label{eightythree}
\end{equation}

\noindent with $\Delta V_{eff}^{T_*}(\varphi_2) = V_{eff}^{T_*}(\varphi_2^{\rm true})-V_{eff}^{T_*}(\varphi_2^{\rm false})$ and $\Delta s$ is the entropy density variation.\\

\noindent Before we move to calculate the phase transition parameters using the above equations, we note that for certain regions of parameter space of our potential in Eq.~(\ref{seventysix}), the decay rate of the false vacuum might be dominated by the $\mathcal{O}(4)$ symmetric Euclidean action $S_4$ instead of the $\mathcal{O}(3)$ symmetric action $S_3$ and in general one has to consider \cite{Ghoshal:2020vud}:
\begin{equation}
    \Gamma = \max\left\{T^4\left(\frac{S_3}{2\pi T}\right)^{3/2}\exp(-S_3/T),\; \frac{1}{R_4^4}\left(\frac{S_4}{2\pi}\right)^2\exp(-S_4)\right\},
    \label{eightyfour}
\end{equation}
where $\Gamma$ is the decay rate of the false vacuum and $R_4$ is the radius of bubbles at nucleation for the $\mathcal{O}(4)$ bounce.\\

\noindent During the $\varphi_2$ phase transition, collisions of expanding false-vacuum bubbles can result in production of $\varphi_1$ particles. The produced density will be non-thermal as $m_{\varphi_1}\sim \sqrt{2\lambda_1}v_1\gg v_2\sim T_*$ which can
potentially source leptogenesis alongside a characteristic stochastic GW signal. In this work, we realize this mechanism using the scalar-portal setup of Ref.~\cite{cataldi2024leptogenesisbubblecollisions}, unlike the RHN-portal framework employed in Secs.~\ref{IV} and~\ref{V}. In the scalar-portal setup, an initial density of the scalar $\varphi_1$ particles is produced from the $\varphi_2$ bubble collisions which subsequently decays into the RHNs.\\ 

\noindent As discussed in Sec.~\ref{IV}, for the particle production formalism to remain valid, the physical masses of the produced particles must not be dominated by the VEV contribution coming from the phase transition\footnote{See the discussion below Fig.~\ref{fig:4} for details.}. 
 However, in the multi-Majoron model under consideration, the right-handed neutrinos (RHNs) predominantly acquire their masses from the same symmetry-breaking phase transition. Moreover, due to the $U(1)_{B-L}\otimes U(1)_N$ symmetry, an additional tree-level Majorana mass term of the form $M_N\overline{N^c}N$ cannot be consistently introduced in the Lagrangian. As a result, the particle-production formalism outlined in Sec.~\ref{II} is no longer applicable when RHNs are produced directly during the phase transition.\\

\begin{figure}[H]
    \centering
    \includegraphics[width=0.6\linewidth]{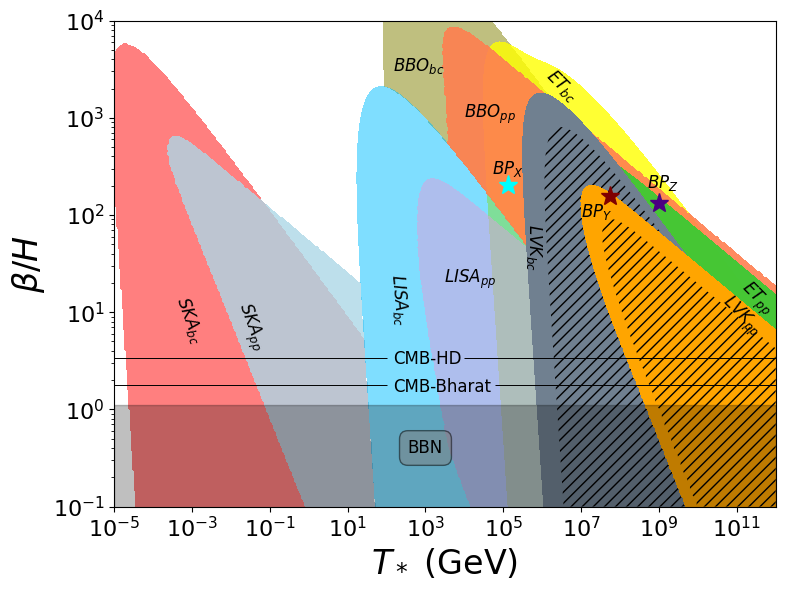}
    \caption{\it Three Benchmark points $BP_X$, $BP_Y$ and $BP_Z$ for different values of $\lambda_2$, $\zeta$, $M$ and $v_2$. The shaded regions corresponds to SNR $>10$ for various GW experiments. The hatched region is ruled out by current LVK observations. $\alpha = 400$ and $\kappa_{\varphi 1} \sim \mathcal{O}(1)$ taken for $\varphi_1$ production.}
    \label{fig:fourteen}
\end{figure}

\noindent In contrast, this issue is avoided if scalar $\varphi_1$ quanta are produced during the $\varphi_2$ phase transition via the mixing interaction proportional to $\zeta$. This is ensured by the hierarchy between the symmetry-breaking scales, $v_1 \gg v_2$, which guarantees that the physical mass of $\varphi_1$ is largely insensitive to the dynamics of the $\varphi_2$ transition. The nonthermally produced $\varphi_1$ particles subsequently decay into the heaviest RHN, $N_3$, through the Yukawa interaction with coupling $y_3$. The decay of $N_3$ into Standard Model states then proceeds asymmetrically, generating the observed Baryon asymmetry. The GW efficiency factor associated with $\varphi_1$ production is obtained from Eq.~(\ref{fiftyone}),
\begin{equation}
    \kappa_{\varphi1} \approx \frac{3}{64\pi^5}\frac{\zeta^2}{\lambda}\ln[\dots]\left(1+\mathcal{O}\left(M_{pl}^{-1}\right)\right).
\end{equation}
 Here $\lambda$ is related to the released vacuum energy for the $\varphi_2$ phase transition $\Delta V \sim \lambda v_2^4$. Having these considerations in mind, we next identify the parameter space for leptogenesis along with its corresponding GW signals. This model has only  three parameters relevant for the GW signal: $(\lambda_2,v_2,\zeta)$. We quote below certain choices of these parameters as the benchmark points and plot them on $\beta/H$ vs $T_*$ plane along with GW SNRs in Fig.~\ref{fig:fourteen}.

 \noindent Next we calculate the phase transition parameters $\alpha, \ \beta/H$ and $T_*$ from the potential in Eq.~(\ref{seventyeigth}) by evaluating the Euclidean action $S_3$ and using the relations Eq.~(\ref{eightyone}-\ref{eightythree}). By performing a random parameter scan of the variables $\{\lambda_2,v_2,\zeta,M\}$ we note down three benchmark points\footnote{We have checked  explicitly for the benchmark points $S_4\gg S_3/T$. Using Eq.~(\ref{eightyfour}) we then take $S_E = S_3/T$ for bounce calculation.} $BP_X$, $BP_Y$ and $BP_Z$ in Fig.~\ref{fig:fourteen}. Corresponding values of different coupling strengths, masses and P.T. parameters are summarized in Table.~\ref{tab-two}.

\begin{table}[H] \centering
\begin{tabular}{|c||c|c|c|c|c|c|c|c|}
\hline
& $\lambda_2$&$v_2$(TeV)& $\zeta$& M (TeV) &$\alpha$&$\beta/H$&$T_*$(TeV)&$\langle\varphi_2\rangle_{T_*}^{\rm true}$ (TeV)\\ \hline\hline
$BP_X$& $1.46\times 10^{-5}$&0.23&0.0460& 0.015&321.7& 204.0& $1.3\times10^2$& $5.6\times10^4$\\
$BP_Y$& $2.8\times10^{-6}$&$75.7$&0.0332&2.4&342.6&156.7 &$5.5\times10^4$&$2.8\times 10^7$ \\
$BP_Z$& $2.3\times10^{-5}$&$9604.5$&$0.0259$&765.2&499.9&134.2&$1.01\times10^6$ &$6.69\times 10^8$\\ \hline
\end{tabular}
\caption{\it Parameter values for the chosen Benchmark points with $\lambda_1 = 0.001$ and $ v_1 = 10^{14}$ GeV.}
\label{tab-two}
\end{table}

\begin{table}[H]
\centering
\begin{tabular}{|l||c|c|c|c|c|c|c|}
\hline
Benchmark & \multicolumn{2}{|c|}{Dark Matter} & \multicolumn{2}{|c|}{Leptogenesis} & \multicolumn{3}{|c|}{Co-genesis} \\
\hline
& $M_3$(TeV) & $\Omega_{DM}h^2$ & $M_3$(TeV) & $\mathcal{Y}_B$ & $M_3$(TeV) & $\mathcal{Y}_B/\text{Br}_L$&$\text{Br}_{\chi}m_\chi$(GeV) \\
\hline\hline
$BP_X$ & $1.1\times10^7$ & 0.122& $2.24\times 10^9$ & $4.08\times10^{-11}$ & $2.24\times 10^9$ & $4.08\times10^{-11}$ & 45.9\\
\hline
$BP_Y$ & $1.55\times10^6$ & 0.121 & $1.17\times 10^8$ & $8.37\times10^{-11}$ & $2.5\times 10^8$ & $3.72\times10^{-10}$ & 5.0\\
\hline
$BP_Z$ & $5.68\times10^5$ & 0.121 & $2.65\times10^{7}$ & $8.75\times10^{-11}$& $3.0\times10^{7}$ & $1.12\times10^{-10}$ & $16.7$\\
\hline
\end{tabular}
\caption{\it BAU and DM relic for the given benchmarks for $\varphi_1$ production and subsequent decays to RHNs. }
\label{tab-3}
\end{table}

\noindent Fig.~\ref{fig:fourteen} shows for all three Benchmarks, the GWs produced from the corresponding phase transition can be probed in detectors BBO and ET with high SNR. This will be possible for GWs coming from the novel particle production mechanism alongside with the well established bubble collision GWs \footnote{Besides there can also be GW arising due to global cosmic strings from global $U(1)_{\rm B-L}$ but those are hardly in the detectable range for the choice of $v_1$ and moreover the estimates for GW from global cosmic strings have huge uncertainty and may vary orders of magnitude, so we do not discuss them in this analysis, see Refs. \cite{Fu:2023nrn,Fu:2025dlp} for some detail.}. As was shown in Fig.~\ref{fig:1}, the RHN production sources GW signals which have a unique characteristic tail of $\Omega_{GW}h^2 \sim f^1$ for lower frequencies and hence can be separated from the GW signals coming from bubble collisions which has the form approximate $\Omega_{GW}h^2 \sim f^{2.3}$.\\

\noindent For the benchmark point $BP_X$, we find that ET is sensitive only to the GW spectrum sourced by bubble collisions, while BBO can probe both the bubble-collision signal and the GW spectrum arising from $\varphi_1$ production. In contrast, for $BP_Z$, BBO is sensitive exclusively to the bubble-collision GW spectrum, whereas ET can probe both GW contributions. For $BP_Y$, although both ET and BBO are, in principle, sensitive to a GW signal, this region of parameter space is already excluded by current LVK observations. Consequently, for the viable benchmark points $BP_X$ and $BP_Z$, a complementary detection pattern with observation of the bubble-collision GW signal in ET followed by the $\varphi_1$-production GW signal in BBO, or vice versa would provide a strong indication of $\varphi_1$ generation during a first-order phase transition.\\ 

For each benchmark point we also calculate the corresponding BAU and DM relics for the three scenarios discussed in sec~\ref{IV}\&\ref{V}. The yield of the produced $N_3$ number density in the scalar portal setup is given as (see Appendix-\ref{app-A} and Ref.~\cite{cataldi2024leptogenesisbubblecollisions} ) 
\begin{multline}
    \qquad\mathcal{Y}_N = \rm{Br}_{\varphi_1\rightarrow N_3}\mathcal{Y}_{\varphi_1}\\ \approx \frac{1}{8\pi^2}\frac{\zeta^2 y_{3}^2}{\zeta^2+y_{3}^2}\left(\frac{\beta}{H}\right)\left(\frac{\pi^2\alpha}{30(1+\alpha)g_*\lambda}\right)^{1/4}\frac{\langle\varphi_2\rangle^{\rm true}_{T_*}}{M_{pl}}\left[\ln\left(\frac{E_{\max}}{m_{\varphi_1}}\right)+\frac{8\pi^2v_1^2}{m_{\varphi_1}^2}\right],\quad
\end{multline}

where $\rm{Br}_{\varphi_1\rightarrow N_3}$ is the branching ratio of $\varphi_1$ into the RHN $N_3$ state, and $E_{\max}$ is given in Eq.~(\ref{45}). Note due to the mixing between $\phi_1$ and $\phi_2$, we have to diagonalize the mass matrix coming from Eq.~\ref{seventysix} in order to find $m_{\varphi_1}$. Following Ref.~\cite{DiBari:2023mwu} we find the corrected mass 
\begin{equation}
    m_{\varphi_1}^2 \approx 2\lambda_1 v_1^2 + \frac{\zeta^2}{2\lambda_1}v_2^2 +\mathcal{O}\left(\left(\frac{v_2}{v_1}\right)^3\right).
\end{equation}
Note after $\varphi_1$ P.T., $N_3$ gets a mass $M_3 = y_{3}v_1/2$. Hence in order for $\varphi_1$ to decay into a pair of $N_3$ one requires
\begin{equation}
    m_{\varphi_1}\gtrsim 2M_3 \implies y_{3}^2\lesssim 2\lambda_1\left(1-\frac{\zeta^2}{4\lambda_1^2}\left(\frac{v_2}{v_1}\right)^2\right).
    \label{82}
\end{equation}
Hence for our parameter choice with $v_1 = 10^{14}$ GeV, we can neglect the vev correction and simply get an upper bound $y_{3}\lesssim 0.45$. This leads to a corresponding upper bound on $M_3$ while calculating different relic density and yields. Note as was discussed earlier, the two lighter RHNs are naturally quasi-degenerate whereas the heaviest RHN exhibits a hierarchical mass scale $M_3\gg M_{1,2}$. This near degenerate masses of the RHNs then can increase the CP violation in the framework of resonant leptogenesis~\cite{Pilaftsis:2003gt}. In particular, $\epsilon_{\rm max}\sim\mathcal{O}(1)$ can be taken if $|M_1-M_2| \sim \Gamma_{1,2}/2$, where $\Gamma_{1,2}$ is the decay rate of the RHNs $N_1,N_2$. This can be achieved by adjusting the corresponding Yukawa couplings $y_1$ and $y_2$ and hence for the rest of this section we will be taking $\epsilon_{CP} \sim \mathcal{O}(1)$.

 Note, in our case $\lambda_2/\zeta \lesssim \mathcal{O}(10^{-3})$ hence we can safely assume $\varphi_2$ self production from $\varphi_2$ bubble collision remains subdominant to $\varphi_1$ production. By inspecting Eq.~\ref{46} we see the produced $N_3$ yield from $\varphi_2$ collision is $\propto \ln\left(\frac{2 \rm E_{max}}{M_3}\right)$ whereas the produced $\varphi_2$ yield is $\propto \left[\ln\left(\frac{ \rm E_{max}}{m_{\varphi_1}}\right) + \frac{8\pi^2v_1^2}{m_{\varphi_1}^2}\right]$. We note for our benchmark points, $\frac{8\pi^2v_1^2}{m_{\varphi_1}^2}\gg \ln\left(\frac{ \rm E_{max}}{m_{\varphi_1}}\right),\ln\left(\frac{2 \rm E_{max}}{M_3}\right)$, hence the produced $N_3$ density always remain subdominant to the produced $\varphi_1$ density. Finally, we point out that the mass of the produced $\varphi_1$ particles, $m_{\varphi_1}\sim \sqrt{2\lambda_1}v_1 \gg v_2 \sim T_*$ for our benchmark and hence the thermal contribution for $\varphi_1$ yield is negligible at production.

 The corresponding values of the DM relic and BAU for the benchmark points are summarized in Table.~\ref{tab-3}.Using Eqs.~(\ref{fortyfive}) and (\ref{48}), we find that in the RHN dark matter scenario the observed relic abundance is reproduced for RHN masses in the range $M_3\sim10^5$–$10^7$~TeV. For leptogenesis, the baryon yield $\mathcal{Y}_B$ is computed using Eq.~(\ref{sixtyfour}), assuming $\epsilon_{\rm max}\sim\mathcal{O}(1)$~\cite{Pilaftsis:2003gt}. For $BP_Y$ and $BP_Z$ we produce the observed baryon yield given in Eq.~\ref{52} for RHN mass $10^7-10^8$ TeV. On the other hand, for $BP_X$, even for the maximum $M_3$ allowed by the bound Eq.~\ref{82}, roughly half of the required asymmetry is produced. This happens due to the relatively low P.T. temperature for $BP_X$.

 The model also allows for a co-genesis scenario through the inclusion of an additional interaction term of the form $\mathcal{L}\sim y_\chi \bar{\chi}\phi_2 N_3$ in Eq.~\ref{seventyfive}. The additional dark sector fermion $\chi$ is charged under $U(1)_N \times U(1)_{B-L}$ such that this operator is invariant under the extended multi-Majoron symmetry. The fermion $\chi$ then serves as the DM candidate. The analysis closely parallels that presented in Sec.~\ref{V}. The inclusion of this extra fermionic degree of freedom does not significantly affect the effective thermal potential, since both the thermal and one-loop corrections from fermions are relatively small~\cite{Salvio:2024upo}. Consequently, the phase transition parameters may be taken to be identical to those computed previously.

In the co-genesis scenario, the dark matter relic abundance is reproduced for $\text{Br}\chi m\chi \sim 5$–$50$~GeV. For $\text{Br}_\chi\sim1$, this corresponds to heavy dark matter with $m_\chi\sim10$~GeV; however, in this limit $\text{Br}_L\ll1$ and the resulting baryon asymmetry is negligible. Conversely, for $\text{Br}_\chi\sim0.1$, the observed baryon asymmetry can be obtained simultaneously with a heavy dark matter candidate, $m_\chi\sim\mathcal{O}(100)$~GeV.\\

\noindent \noindent Before concluding this section, we perform a parameter scan of the phase transition in the $v_2-M$ plane and identify regions consistent with successful dark matter production, leptogenesis, and co-genesis. On the same plane, we also display the gravitational-wave SNRs for the ET and BBO detectors, for GW signals originating from both bubble collisions and $\varphi_1$ production.

\begin{figure}[H]
    \centering
    \includegraphics[width=0.49\linewidth]{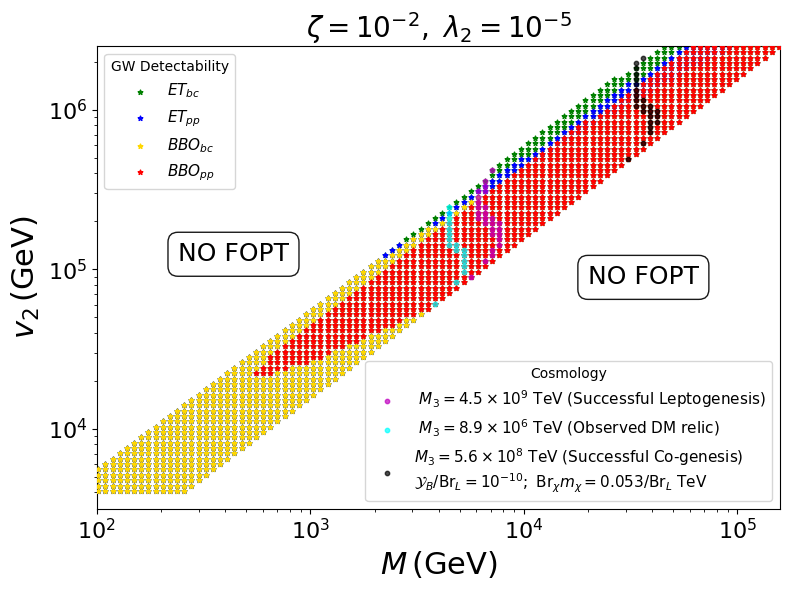}
    \includegraphics[width = 0.49\linewidth]{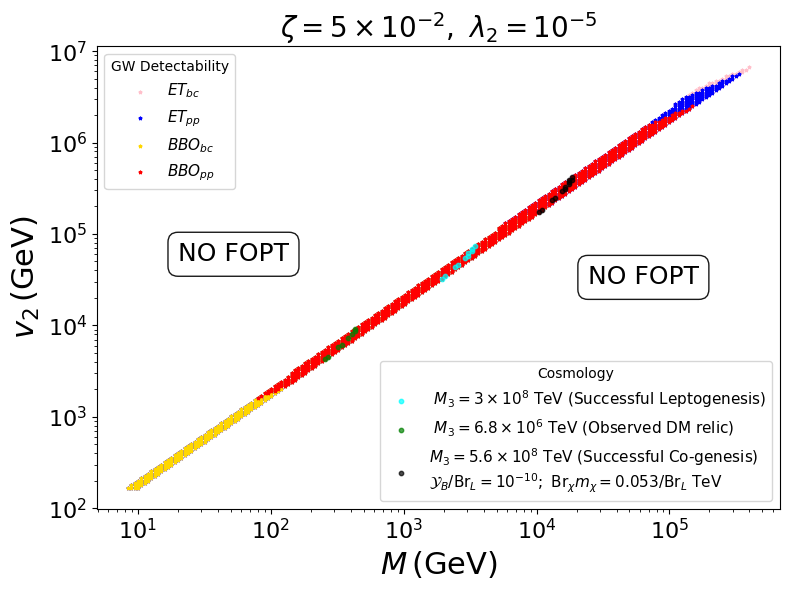}
    \includegraphics[width = 0.49\linewidth]{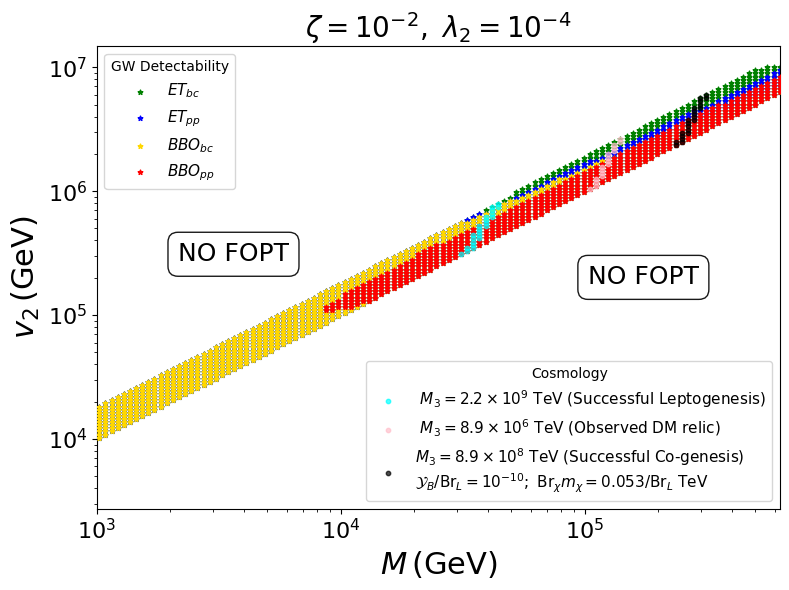}
    \caption{\it Parameter scan for a FOPT in the multi-Majoron model in $v_2-M$ plane for differnt choices of $\zeta$ and $\lambda_2$. Shown are parameter-space points that realize a successful FOPT and produce GW signals with SNR $>10$ at ET and BBO, arising from both bubble collisions and $\varphi_1$ production. Regions compatible with successful dark matter production, leptogenesis, and co-genesis are indicated, together with the corresponding values of the RHN mass $M_3$. $\kappa_{\varphi_1}\sim\lambda\sim\mathcal{O}(1)$, $v_1 = 10^{14}$ GeV and $\lambda_1 = 0.001$ is taken for all the plots.}
    \label{fig:sixteen}
\end{figure}

\noindent Figure~\ref{fig:sixteen} demonstrates that, over a wide region of parameter space, the multi-Majoron model can accommodate first-order phase transitions (FOPTs) that are detectable with high signal-to-noise ratio (SNR) at the future gravitational-wave experiments ET and BBO. In particular, for $M\gtrsim 10^4$ GeV, varying $v_2$ allows for FOPTs that can be probed through the novel GW signal arising from $\varphi_1$ production at BBO, together with the corresponding bubble-collision GW signal detectable at ET. Note, at any given $v_2$ with fixed $\zeta$, $\lambda_1$ and $\lambda_2$, only free parameter left is $y_{1}\sim y_2$. Hence FOPT is achieved for only a narrow band of the Yukawa couplings at a given $v_2$, leading to small variation in $M$. As $v_2$ increases, since $M \sim y_{1}v_2$, we see an increase in $M$ with the same narrow span. For the same value of $\zeta$, increasing $\lambda_2$ increases the scale of P.T. and RHN mass $M$. On the other hand, for the same value of $\lambda_2$, increasing $\zeta$ results in less amount of successful FOPT points.

For $v_2\in[10^5,10^6]$ GeV and $M\in [3\times10^3,4\times10^4]$ GeV,the parameter space admits regions where all three scenarios: dark matter, leptogenesis, and co-genesis can be realized separately. In the RHN dark matter case, the model predicts ultraheavy dark matter with $ m_{DM}\equiv M_3 = 8.9\times10^6$ TeV. Successful high-scale leptogenesis is achieved for $M_3 = 4.5\times 10^9$ TeV. In addition, a high-scale co-genesis scenario is possible, in which the dark matter candidate acquires a mass of $\sim \mathcal{O}(100) \ \rm GeV$.

The corresponding GW signals in all three scenarios are detectable with high SNR and can be distinguished through complementary observations by ET and BBO. This establishes a concrete proof of concept where capitalizing on the novel particle-production GW signals, one can detect BSM physics in a UV complete model.

\medskip

\section{Discussion \& Conclusion}\label{conclusion}

During a first-order phase transition (FOPT), particles produced in bubble–wall collisions generate a novel contribution to the stochastic gravitational-wave background. If right-handed neutrinos (RHNs) are produced through this mechanism, the resulting population can account for the observed baryon asymmetry via leptogenesis when they decay. They may constitute the present DM relic abundance in the case of the RHNs are stable. Finally if the RHNs decay into a dark sector, it may simultaneously account of DM and BAU, explaining the co-incidence $\Omega_{\rm DM} \sim 5 \Omega_{\rm B}$ through co-genesis. The associated characteristic gravitational-wave (GW) signatures from RHN production provide a complementary probe of these scenarios. An overview of these scenarios is presented in Fig.~\ref{fig:diag}. The key findings of our analysis are summarized below:

\begin{itemize}

\item We showed the stochastic GW background from both bubble collisions and particle production can be probed by several current and future GW detectors, including SKA, LISA, ET, BBO, and certain regions of parameter space has already been ruled out by LVK $\mathcal{O}(3)$ data (see Fig.~\ref{fig:1}). The particle production from bubble wall introduces a characteristic low-frequency tail, $\Omega_{\rm GW} h^2 \propto f^{1}$ different from the bubble-collision tail which is approximately $\Omega_{\rm GW} h^2 \propto f^{2.3}$. To illustrate the complementarity among the GW detectors, we investigated the parameter space of phase transitions in the $(\beta/H,,T_*)$ plane (see Fig.~\ref{fig:2}). For instance, for a FOPT with $\beta/H = 10$ and $T_{*} = 100$ GeV, the particle production GW signal is detectable in SKA while the bubble-collision GWs lies within LISA’s reach.

\item A non-thermal population of RHNs can be generated from bubble collisions during a FOPT. If the lightest among them is stable then they may constitute the entire DM relic abundance for masses as low as $m_{DM}\gtrsim 10^{6}\ \mathrm{GeV}$ (Fig.~\ref{fig:4}). Higher transition temperatures correspond to lighter viable DM masses (see Eqs.~\ref{fortyfive} and \ref{48}).
 The characteristic  $f^{1}$ slope from RHN production will be detectable in LISA for $m_{DM}\sim 10^{10}\ \mathrm{GeV}$, SKA for $m_{DM}\sim 10^{14}\ \mathrm{GeV}$ and ET, BBO and future LVK detectors for $m_{DM}\sim 10^{6}\ \mathrm{GeV}$ (see Fig.~\ref{fig:5}). The novel correlation between the DM relic, mass and the GW amplitude can be understood  from Eqs.~(\ref{30}-\ref{38},\ref{46}-\ref{48}) via $\alpha,\ \beta/H,\ T_*$ and $y_{\rm DM}$.

\item Once RHNs are produced from the cosmic bubble collisions, non-thermal leptogenesis is achieved through a Yukawa coupling $\lambda_N$ between light and heavy neutrinos (see Eq.~\ref{fiftytwo}). Observed BAU is obtained for $M_1 \gtrsim 10^{11}\ \mathrm{GeV}$ and $T_*\gtrsim 10^{6},\mathrm{GeV}$ (Fig.~\ref{fig:8}, \ref{fig:nin3}). For smaller $M_1$, a higher $T_*$ is required because the asymmetry generated scales as $Y_B\propto T_*\cdot M_1\cdot\ln{(1/M_1)}$ (see Eqs.~\ref{51},\ref{BAU}). A significant part of the parameter space is already constrained by LVK (Fig.~\ref{fig:9}) from non-observation of SGWB. The GWs arising due to the RHN production in the allowed ranges are testable in upcoming ET and BBO detectors. As an example, a FOPT with $\beta/H = 150$ and $T_* = 10^{9}$ GeV yields successful leptogenesis while producing bubble-collision GWs detectable in ET and future LVK with RHN-production GWs detectable in BBO, enabling inter-detector complementarity. The correlation between the novel GW source from RHN production and the leptogenesis microphysics can be seen in Eqs.~(\ref{30}-\ref{38},\ref{CP}-\ref{BAU}).

\item If RHNs decay into a dark sector state $\chi$ through a CP-violating coupling (see Eq.~\ref{sixtyeight}), their decay products can account for the present DM abundance via asymmetric dark matter with $m_{\rm DM}\in[10^{-4},10^{4}],\mathrm{GeV}$, while simultaneously generating the observed baryon asymmetry. Successful co-genesis occurs for $T_* \gtrsim 10^{7},\mathrm{GeV}$ and $M_1 \gtrsim 10^{9},\mathrm{GeV}$. The resulting GW signal from RHN production, together with bubble-collision GWs, lies within the reach of LISA, BBO, and ET (see Fig.~\ref{fig:13}). For instance, having successful co-genesis, for $\beta/H = 150$, ET can probe both components for $T_* \in [10^{7},10^{10}]$ GeV ($m_{\rm DM} \gtrsim 1$ GeV), while BBO is sensitive only to the RHN-production component for $T_* \in [10^{7},10^{8}]$ GeV ($m_{\rm DM} \gtrsim 100$ GeV). Thus, in this regime, a bubble-collision detection in ET together with an RHN-production detection in BBO would provide a clean inter-detector test of the co-genesis scenario. The dependencies of the novel GW signal on the co-genesis parameters can be understood from Eqs.~(\ref{30}-\ref{38},\ref{69}-\ref{73}) where the cosmic co-incidence of $\Omega_{\rm CDM} \sim 5 \Omega_{\rm b}$ is dynamically explained from RHN decay.

\item Finally, in Sec.~\ref{VI}, we calculate the phase-transition parameters $\alpha$, $\beta/H$, and $T_*$ in a UV-complete multi-Majoron model with a global $ U(1)_N \times U(1)_{B-L}$ extension of the SM (Table~\ref{tab-two}), motivated to explain the hierarchy of neutrino masses, and analyze the resulting GW signatures in Fig.~\ref{fig:sixteen}. Within this \emph{Cosmic Majoron Collider}, we identify regions of parameter space that support strong FOPTs detectable with high SNRs at ET and BBO. In particular, for $M \gtrsim 10^4$ GeV, varying $v_2$ enables FOPTs that can be probed through a novel GW signal from $\varphi_1$ production at BBO, accompanied by a bubble-collision GW signal detectable at ET. Since the vev $v_2$ also sets the seesaw scale, these GW signals provide a direct probe of SM neutrino mass generation.

\item For each parameter point analyzed in Sec.~\ref{VI}, we investigate three distinct cosmological realizations: RHN dark matter, leptogenesis, and co-genesis. We find that for $v_2 \in [10^5,10^6]$ GeV and $M \in [3\times10^3,\,4\times10^4]$ GeV, the parameter space admits regions where each scenario can be realized individually. This constitutes the \emph{first demonstration} that dark matter, leptogenesis, and co-genesis can be accommodated within a unified multi-Majoron framework (see Table~\ref{tab-3}). Consequently, a combined detection of bubble-collision GWs at ET and $\varphi$-induced $N_3$ production GWs at BBO or vice versa would constitute compelling evidence for particle production during a first-order phase transition, simultaneously tracking the seesaw scale and leptogenesis in the multi-Majoron model.

\end{itemize}

\noindent Our analysis demonstrates that within the type-I seesaw framework, first-order phase transitions accompanied by right-handed neutrino production from a \emph{cosmic collider} can simultaneously account for the observed baryon asymmetry and dark matter abundance, while generating distinctive multi-component gravitational-wave spectra. The simultaneous presence of bubble-collision and RHN-production signals, each with characteristic spectral features and potentially detectable across multiple experiments, enables a powerful multi-detector strategy to test scenarios of dark-matter genesis, leptogenesis, and co-genesis. In the context of bubble collisions acting as a \emph{Cosmic Majoron Collider}, FOPTs can access energy scales far exceeding the plasma temperature and potentially approaching the Planck scale, thereby probing new physics coupled to right-handed neutrinos beyond the reach of terrestrial experiments through their primordial GW imprints. As next-generation GW observatories such as LISA and ET come online, the mechanisms explored in this work provide a realistic pathway to probing the microscopic origin of the baryon asymmetry and dark matter, with the potential to transform our understanding of early-Universe physics.

\section*{Acknowledgement}
Authors are grateful to Bibhushan Shakya for communication regarding clarification of several doubts. Authors thank Marek Lewicki, Henda Mansour, Narendra Sahu, Angus Spalding, Luigi delle Rose, Satyabrata Datta and Pankaj Borah for discussions and comments on the manuscript. Authors thank the \href{https://www.icts.res.in/}{International Centre for Theoretical Sciences (ICTS)} Bangalore, India for the arrangement of workshop \href{https://www.icts.res.in/program/gwbsm2024} {"Hearing beyond the standard model with cosmic sources of Gravitational Waves"}  where the project got initiated.

\appendix
\section{Theoretical Framework for Gravitational Waves}
\label{app-A}

\noindent For a homogeneous isotropic universe the background metric is given by the Friedmann –Robertson–Walker (FRW) metric. In presence of tiny inhomogeneities the metric picks up perturbations which are perceived as gravitational waves 
\begin{align}
    &ds^2 = -dt^2 + a^2(t)\left[\delta_{ij} + h_{ij}(\Vec{x},t)\right]d\Vec{x}^2.\\
    &h_{ij} (\Vec{x},t) = \frac{1}{(2\pi)^3} \int d^3q e^{-i\Vec{q}\cdot\Vec{x}}h_{ij} (\Vec{q},t).
\end{align}

\noindent Here $a(t)$ is the scale factor of the universe and $t$ is the cosmic time in the FRW metric. These perturbations are sourced by the transverse traceless part of the energy-momentum tensor ($T_{ij}$) of the matter distribution and satisfy conditions $h_{ii} = \partial_j h_{ij} = 0$ due to transverse and traceless conditions. We assume that the phase transition completes within a timescale significantly smaller than Hubble time ($t_*\ll H^{-1}(t_*)$, where $t_*$ denotes phase transition time), so that the expansion of space can be neglected. Writing the Einstein equation and going to the transverse traceless gauge we find
\begin{equation}
    \ddot{h}_{ij} (\Vec{q},t) + q^2h_{ij} (\Vec{q},t) = 8\pi G\ \Pi_{ij}^T(\Vec{q},t),
    \label{15}
\end{equation}

\noindent where $\Pi_{ij}^T$ is the transverse traceless part of the energy momentum tensor
\begin{equation*}
    T_{ij}(t,\Vec{x}) = \Pi_{ij}^T (t,\Vec{x}) + ...\quad ; \quad \Pi_{ij}^T (t,\Vec{x}) = \frac{1}{(2\pi)^3} \int d^3q e^{-i\Vec{q}\cdot\Vec{x}}\Pi_{ij}^T (\Vec{q},t)
\end{equation*}
 and $T_{ij}(t,\vec{x})$ is the full energy momentum tensor. The gravitational wave energy density, as denoted by $\rho_{GW}$ is defined as 
\begin{equation}
    \rho_{GW}(t) = \frac{1}{32\pi G}\langle \Dot{h}_{ij}(t,\Vec{x})\Dot{h}_{ij}(t,\Vec{x})\rangle.
\end{equation}
In the Fourier space we define the equal time correlator as :
\begin{equation}
    \langle \Dot{h}_{ij}(t,\Vec{q_1})\Dot{h}^*_{ij}(t,\Vec{q_2})\rangle \equiv (2\pi)^3 \delta^3(\Vec{q}_1-\Vec{q}_2)P_{\Dot{h}}(q_1,t),
    \label{17}
\end{equation} then we find the Gravitational wave energy density\cite{Caprini_2018} 
\begin{align*}
    \rho_{GW}(t) = \frac{1}{32\pi G}\frac{1}{2\pi^2}\int dq q^2 P_{\Dot{h}}(q,t) \implies \frac{d\rho}{d \ln{q}} = \frac{1}{64\pi^3 G}q^3P_{\Dot{h}}(q,t).
\end{align*}

\noindent We define the power spectrum to be $\Omega_{GW} \equiv \frac{1}{\rho_{tot}}\frac{d\rho_{GW}}{d\ln{q}}$ and finally get 
\begin{equation}
    \Omega_{GW}(t,q) = \frac{1}{24\pi^2 H^2}q^3 P_{\Dot{h}}(q,t) .
    \label{18}
\end{equation}
Hence in order to find the GW power spectrum one needs to figure out the Fourier transformed 2-point correlator. We can achieve this by solving Eq.~(\ref{15}) with the Greens function. Assuming the source term is active from $t_{start} \rightarrow t_{end}$, let 
\begin{equation}
    h_{ij}(\vec{k},t) = 8\pi G\int_{t_{start}}^{t_{end}} dt' G_k(t,t')\Pi_{ij}(t',\vec{k}).
    \label{19}
\end{equation}
Then the Greens Function $G(t,t')$ solves the homogeneous part of Eq.~(\ref{15}) such that 
\begin{equation}
    \frac{\partial^2G_k(t,t')}{\partial t^2} + k^2 G_k(t,t') = \delta(t-t').
\end{equation}
The solution to this is found with the boundary conditions : $G_k(t,t) = 0\ ;\ \left. \frac{\partial G_k}{\partial t}\right|_{t = t'} = 1$, which is given by : $G_k(t,t') = \frac{\sin{(k(t-t'))}}{k}$. Substituting this in Eq.~(\ref{19}) we find 
\begin{equation}
    h_{ij}(t,\vec{k}) = A_{ij}(\vec{k})\sin\left(k(t-t_e)\right) + B_{ij}(\vec{k})\cos(k(t-t_e)),
    \label{21}
\end{equation}
where, $t_e = t_{end}$ , $t_s = t_{start}$ and 
\begin{align}
    A_{ij}(\vec{k}) &= \frac{8\pi G}{k}\int_{t_s}^{t_e} dt \cos\left(k(t_e-t)\right)\Pi_{ij}(t,\vec{k})\ ; \\
    B_{ij}(\vec{k}) &= \frac{8\pi G}{k}\int_{t_s}^{t_e} dt \sin\left(k(t_e-t)\right)\Pi_{ij}(t,\vec{k}).
\end{align}
Now in Eq.~(\ref{17},\ref{18}) we see the power spectrum is dependent on the equal time correlator of the metric perturbations $h_{ij}$'s. Substituting Eq.~(\ref{21}) in Eq.~(\ref{17}) we see the power spectrum is dependent on the Unequal Time Correlator (UETC) of the transverse energy momentum tensor $\Pi_{ij}$'s of the source . We define 
\begin{equation}
    \left\langle\Pi_{ij}(t_x,\vec{k})\Pi_{ij}^*(t_y,\vec{q})\right\rangle \equiv (2\pi)^3\delta^{(3)}(\vec{k}-\vec{q})\Pi(t_x,t_y,k) = \Lambda_{ij,kl}(\hat{k})\Lambda_{ij,mn}(\hat{q}) \left\langle T_{kl}(t_x,\vec{k})T_{mn}(t_y,\vec{q})\right\rangle 
    \label{24}
\end{equation}
where $\Lambda_{ij,kl}(\hat{k})$ is the transverse traceless projection tensor in the $\hat{k}$ direction. From Eq.~(\ref{19}), using Liebnitz rule of integration we find :
\begin{equation}
    \dot{h}_{ij}(t,\vec{k}) = 8\pi G\int_{t_s}^{t_e} dt' \frac{\partial G(t,t')}{\partial t}\Pi_{ij}(t',\vec{k}).
\end{equation}
Substituting this back in Eq.~(\ref{17}) we get:
\begin{align*}
    (2\pi)^3\delta^{(3)}(\vec{k}-\vec{q}) P_{\dot{h}}(t,\vec{k}) &= (8\pi G)^2 \int_{t_s}^{t}dt_x\int_{t_s}^{t}dt_y\frac{\partial G_k}{\partial t}(t,t_x)\frac{\partial G_q^*}{\partial t}(t,t_y)\left\langle\Pi_{ij}(t_x,\vec{k})\Pi_{ij}^*(t_y,\vec{q})\right\rangle\\
    \implies P_{\dot{h}}(t,\vec{k}) &=64\pi^2G^2 \int_{t_s}^{t}dt_x\int_{t_s}^{t}dt_y \cos{[k(t-t_x)]}\cos{[k(t-t_y)]}\Pi(t_x,t_y,k)\\
    &= 32\pi^2G^2\int_{t_s}^{t}dt_x\int_{t_s}^{t}dt_y\cos{[k(t_x-t_y)]}\Pi(t_x,t_y,k).
\end{align*}
Here,
\begin{equation*}
    \Pi(t_x,t_y,k) = (2\pi)^3\delta^{(3)}(\vec{k}-\vec{q})\left\langle\Pi_{ij}(t_x,\vec{k})\Pi_{ij}^*(t_y,\vec{q})\right\rangle
\end{equation*}
is the UETC. Putting this back in Eq.~(\ref{18}) and adjusting the factors we get :
\begin{equation}
    \Omega_{GW}(t,k) \equiv \frac{1}{\rho_{tot}}\frac{d\rho_{GW}}{d\ln{k}} = \frac{2G k^3}{\pi\rho_{tot}} \int_{t_s}^{t}dt_x\int_{t_s}^{t}dt_y\cos{[k(t_x-t_y)]}\Pi(t_x,t_y,k).
    \label{26}
\end{equation}
This is the final formula for the GW power spectrum. We see for different sources all we have to calculate is the UETC of the transverse traceless energy momentum tensor and substitute it back in Eq.~(\ref{26}) to arrive at the GW spectrum.

\section{Particle Production for Scalar and Vector Bosons}
\label{app-B}
The particle production mechanism can not only produce fermionic particles (as discussed in sec~\ref{II}), but also can produce scalar and vector particles as well. In what follows, we discuss the corresponding 2-point 1 PI Green's functions for these interactions - 
\begin{enumerate}
    \item \textbf{Scalar Self Coupling:} The scalar $\phi$ particles themselves can be produced through the background field excitations, via the quartic term $\mathcal{L}_I\supset -\frac{\lambda_\phi}{4!}\phi^4$ in the scalar potential; this gives rise to a two body $\phi^*\rightarrow \phi\phi$ decay process -
    \begin{equation}
        \text{Im}\left[\tilde{\Gamma}^{(2)}(p^2)\right]_{\phi^*\rightarrow\phi\phi} = \frac{\lambda_\phi^2 v_\phi^2}{8\pi}\left(1-\frac{4m_\phi^2}{p^2}\right)\Theta(p-2m_\phi)
    \end{equation}
    And a three body $\phi^*\rightarrow 3\phi$ decay process -
    \begin{equation}
        \text{Im}\left[\tilde{\Gamma}^{(2)}(p^2)\right]_{\phi^*\rightarrow\phi\phi\phi} = \frac{\lambda_\phi^2 p^2}{3072\pi^3}\left(1-\frac{9m_\phi^2}{p^2}\right)\Theta(p-3m_\phi)
    \end{equation}
    Note that the three-body process is suppressed relative to the two-body process by a loop factor, as it involves an additional particle in the final state. However, it scales with $p^2$ rather than $v_\phi^2$ and thus can become increasingly significant at higher momentum transfer. In particular, it can remain operative even in the $v_\phi\rightarrow0 $ limit where the symmetry is unbroken.
    
    \item \textbf{Other Scalar Couplings:} For scalars other than $\phi$ itself, for example a dark scalar $\chi$ which is coupled to $\phi$ via the interaction $\mathcal{L}_I \supset - \frac{1}{2}\lambda_\chi \phi^2\chi^2$ we get two body and three body decay processes just like last time :
    \begin{align}
        \text{Im}\left[\tilde{\Gamma}^{(2)}(p^2)\right]_{\phi^*\rightarrow\chi\chi} &= \frac{\lambda_\chi^2 v_\phi^2}{8\pi}\left(1-\frac{4m_\chi^2}{p^2}\right)\Theta(p-2m_\chi)\\
        \text{Im}\left[\tilde{\Gamma}^{(2)}(p^2)\right]_{\phi^*\rightarrow\phi\chi\chi} &= \frac{\lambda_\chi^2 p^2}{1024\pi^3}\left(1-\frac{(m_\phi+2m_\chi)^2}{p^2}\right)\Theta(p-(m_\phi+2m_\chi))
    \end{align}
\end{enumerate}
The calculation for vector particles and final states involving gauge bosons is more complicated due to the gauge dependence of the formalism and hence is out of the scope of this work.

\medskip
\section{Processes leading to Washout of Asymmetries}
\noindent Here we note down a few washout processes related to our scenario discussed in sec~\ref{IV}. The expressions for all scattering cross sections are approximate and we have dropped the contributions of any thermal masses or subleading logarithmic pieces~\cite{cataldi2024leptogenesisbubblecollisions} -
\begin{enumerate}
    \item \bm{$LH\rightarrow N$} : The inverse decay with the decay rate given by -
    \begin{equation}
        \Gamma_{inv} = \frac{\lambda_N^2 M_N}{8\pi}\exp{-M_N/T_*}
    \end{equation}. Since a heavy RHN is in the final state, we see the rate is Boltzmann suppressed and hence is irrelevant for $M_N>T_*$.
    \item  \bm{$Q_3 L\rightarrow N t$} and \bm{$t L\rightarrow N Q_3$} : Scattering of top quarks via intermediate Higgs with the scattering rate given by - 
    \begin{equation}
        \Gamma_{scatter} \approx n_{th}\frac{3\lambda_N^2 y_t^2}{2\pi M_N}\exp{-M_N/T_*}
    \end{equation}
    
    Where $y_t$ is the top Yukawa coupling and $n_{th} = g_t T_*^3/\pi^2$ is the thermal abundance with $g_t = 2$ for top quark. Again due to a heavy RHN in the final state the rate is Boltzmann suppressed and hence is irrelevant. 

    \item \bm{$NL\rightarrow Q_3 t$} : Scattering via s-channel Higgs. The rate is given as -
    \begin{equation}
        \Gamma_{NL\rightarrow Q_3 t} \approx \left(\frac{T_n}{T}\right)^3 n_{th} \frac{3\lambda_N^2 y_t^2}{4\pi(M_N^2 + 4 E_N T_*)}
    \end{equation}
    Where $T_n$ is the nucleation temperature. Note this process is not Boltzmann suppressed like the last two because of the absence of a heavy RHN in the final state. This process shuts off as soon as the RHNs have decayed and hence will be relevant only before this happens.
    \item  \bm{$LL\rightarrow HH$} and \bm{$LH\rightarrow H\bar{L}$} : $\Delta L = 2$ scattering processes with a heavy RHN as intermediate. The rate is given as -
    \begin{equation}
        \Gamma_{(\Delta L = 2)} \approx n_{th} \frac{\lambda_N^4}{8\pi M_N^2} 
    \end{equation}
    Again this process is also not Boltzmann suppressed but is small for large RHN masses.
\end{enumerate}
Since no interacting thermal plasma existed at the time of RHN production or decay, we don't get additional washout processes that one may expect in the standard thermal leptogenesis scenario.

\medskip

\section{RHNs and Radiation Domination}

\noindent We define -
\begin{equation}
    P_1 \equiv \left.\frac{\Gamma_1}{H(T)}\right|_{T = M_1} = \frac{\tilde{m}_1}{2\times 10^{-3}eV}\ ,
\end{equation}
where $H(T)$ is the Hubble expansion rate during Radiation Domination (RD) and the last line follows from the fact - 
\begin{equation}
    \Gamma_1 = \frac{\tilde{m}_1}{8\pi v^2}M_1^2\ ;\quad H(T) = \frac{1}{\sqrt{3}M_{pl}}\left(\frac{\pi^2}{30}g_*(T)\right)^{1/2}T^2.
\end{equation}
 Note the RHNs decay after decoupling from the plasma, i.e. when $\Gamma_1 \lesssim H(T)$. Any particle species becomes non relativistic when the temperature of the plasma becomes lower than its rest mass, i.e. when - $T \lesssim M_1$. This shows, if $P_1 \ll 1$ i.e. if $\tilde{m}_1\ll 2\times 10^{-3}$ eV then $N_1$ decays after becoming non relativistic. This would imply deviation from the Radiation Domination and standard cosmology. Hence we need to find a bound on $\tilde{m}_1$ to ensure radiation domination at all times. 

 To do this, we first consider the temperature $T_{1}$, after which RHN energy density becomes significant that universe exits radiation domination. We can safely assume if RHN energy density is less than 1\% of the radiation density, then universe still remains in radiation domination:
\begin{equation}
    \frac{\rho_{N_1}}{\rho_R} \simeq 10^{-2} \implies T_{1} \simeq \frac{7}{400}\frac{M_1}{g_*(T_{1})} \simeq 2\times 10^{-4}\ M_1 
    \label{59}
\end{equation}
Where $\rho_R$ is the energy density of the radiation bath that scales as $\sim g_*(T)T^4$, $g_* \sim \mathcal{O}(100)$. We assume RHNs start to decay as soon as they decouple from the plasma, at a Temperature $T_{dec}$ given by -
\begin{equation}
    \Gamma_1 = H(T_{dec}) \implies T_{dec} = 3\times 10^8\ \text{GeV}\sqrt{\frac{\tilde{m}_1}{10^{-6}\ \text{eV}}}\left(\frac{M_1}{10^{10}\ \text{GeV}}\right)\left(\frac{106.25}{g_*(T_{dec})}\right)^{1/4}
    \label{60}
\end{equation}
In deriving Eq.~\ref{60}, we have combined Eq.~\ref{56},\ref{57}. Note if RHNs decouple before universe reaches the temperature $T_1$, then radiation domination will be maintained at all time. In other words -
\begin{equation}
    T_{dec} \gtrsim T_{1} \implies \tilde{m}_1 \gtrsim 4\times 10^{-11} \ \text{eV}  
    \label{64}
\end{equation}

\noindent If the RHNs decay almost immediately after they are produced from bubble collision, then the phase transition temperature $T_*$ must be below $T_{dec}$.  At the time of the decay, this energy density in the RHNs will be converted into relativistic d.o.f via couplings to the SM states, whose temperature will coincide with the phase transition temperature $T_{*}$. If radiation domination is maintained at all times then $T_*$ must be greater or equal to $T_{1}$. Hence we get a bound on phase transition temperature that will satisfy our scenario -
\begin{equation}
    T_{dec}\gtrsim T_* \gtrsim T_{dom} \implies 2\times 10^{-4}\lesssim \frac{T_*}{M_1}\lesssim 30\sqrt{\frac{\tilde{m}_1}{\text{eV}}}
    \label{65}
\end{equation}
Where in deriving the last inequality we have used Eq.~(\ref{59},\ref{60}). Hence to summarize, if the bound in Eq.~(\ref{64}) is violated, then universe will deviate from radiation domination at an intermediate time. On the other hand, if Eq.~(\ref{64}) is satisfied but the bound in Eq.~(\ref{65}) is violated, then radiation domination will be maintained at all times but the RHNs will not immediately decay after being produced from the FOPT. Depending upon the different parameter regions one is interested in, corresponding bounds have to be checked.

\nocite{*}

\bibliographystyle{apsrev4-1}

\bibliography{ref}
\end{document}